%% file: top_eeec_prd.tex
\documentclass[prd,aps,twocolumn,nofootinbib,preprintnumbers,amsmath,superscriptaddress,floatfix]{revtex4-1}

\input{commands.tex}

\newcommand{\lett}{article\xspace}

\begin{document}
	
	\title{A New Paradigm for Precision Top Physics:\\Weighing the Top with Energy Correlators}
	
	\author{Jack Holguin}
	\affiliation{CPHT, CNRS, Ecole polytechnique, IP Paris, F-91128 Palaiseau, France}
	
	\author{Ian Moult}
	\affiliation{Department of Physics, Yale University, New Haven, CT 06511}
	
	\author{Aditya Pathak}
	\affiliation{University of Manchester, School of Physics and Astronomy, Manchester, M13 9PL, United Kingdom}
	
	\author{Massimiliano Procura}
	\affiliation{University of Vienna, Faculty of Physics, Boltzmanngasse 5, A-1090 Vienna, Austria}
	
	\preprint{UWThPh 2021-28}
	
	\begin{abstract}
		Final states in collider experiments are characterized by correlation functions, $\langle \mathcal{E}(\vec n_1) \cdots \mathcal{E}(\vec n_k) \rangle$, of the energy flow operator $ \mathcal{E}(\vec n_i)$. We show that the top quark imprints itself as a peak in the three-point correlator at an angle $\zeta \sim m_t^2/p_T^2$, with $m_t$ the top quark mass and $p_T$ its transverse momentum, providing access to one of the most important parameters of the Standard Model in one of the simplest field theoretical observables. 
		Our analysis provides the first step towards a new paradigm for a precise top mass determination that is, for the first time, highly insensitive to soft physics and underlying event contamination whilst remaining directly calculable from the Standard Model Lagrangian.
	\end{abstract}
%

\maketitle

\section{Introduction}
The Higgs and top quark masses play a central role both in determining the structure of the electroweak vacuum~\cite{Degrassi:2012ry,Buttazzo:2013uya,Andreassen:2014gha}, and in the consistency of precision Standard Model fits~\cite{Baak:2012kk,Baak:2014ora}. Indeed, the near-criticality of the electroweak vacuum may be one of the most important clues from the Large Hadron Collider (LHC) for the nature of beyond the Standard Model physics~\cite{Giudice:2006sn,Buttazzo:2013uya,Khoury:2019yoo,Khoury:2019ajl,Kartvelishvili:2020thd,Giudice:2021viw}. This provides strong motivation for improving the precision of Higgs and top mass measurements.

While the measurement of the Higgs mass is conceptually straightforward both theoretically and experimentally~\cite{ATLAS:2015yey}, this could not be further from the case for the top mass ($m_{t}$). Due to its strongly interacting nature, a field theoretic definition of $m_{t}$, and its relation to experimental measurements, is subtle. In $e^+e^-$ colliders, precision $m_{t}$ measurements can be made from the threshold lineshape~\cite{Fadin:1987wz,Fadin:1988fn,Strassler:1990nw,Beneke:1998rk,Beneke:1999qg,Hoang:2000ib,Hoang:2001mm,Beneke:2015kwa}. However, this approach is not possible at hadron colliders, where, despite the fact that direct extractions have measured $m_{t}$ to a remarkable accuracy~\cite{CDF:2014upy,CMS:2015lbj,ATLAS:2016muw,ParticleDataGroup:2020ssz}, there is a  debate on the theoretical interpretation of the measured ``Monte Carlo (MC) top mass parameter''~\cite{Hoang:2008xm}. This has been argued to induce an additional ${\cal O}$(1 GeV) theory uncertainty on $m_t$. For recent discussions, see~\cite{Nason:2017cxd,Hoang:2020iah}. It is therefore crucial to explore kinematic top-mass sensitive observables at the LHC where a direct comparison of the experimental data with first principles theory predictions can be carried out. 

\begin{figure}
	\includegraphics[width=0.32\textwidth]{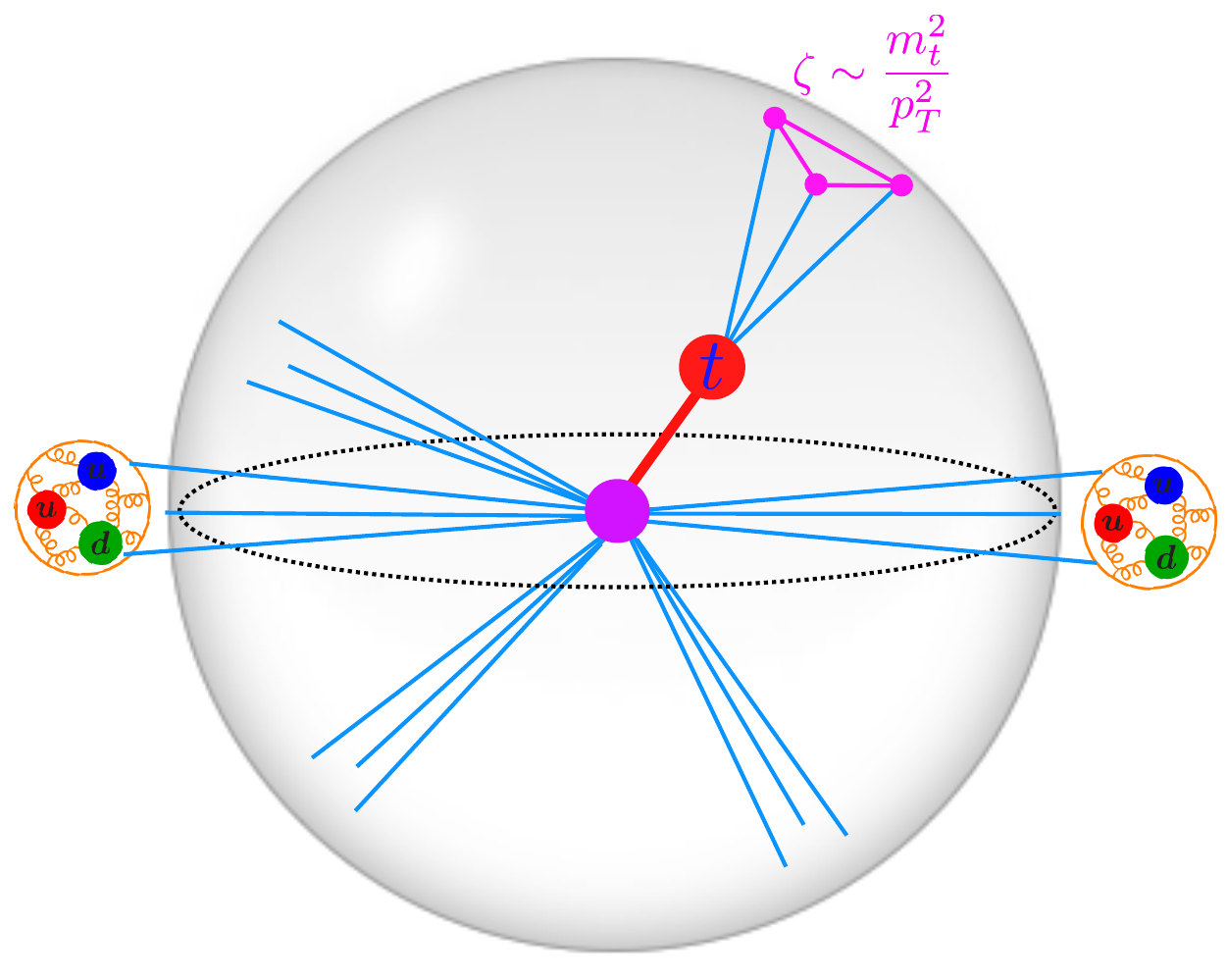}
	\caption{A boosted top quark imprints its short lived existence onto the three-point correlator with a characteristic angle, $\zeta \sim (1- \cos \theta )/2 \sim m_t^2/p_T^2$.}
	\label{fig:top_correlator}
\end{figure}

Significant progress has been made in this regard from multiple directions. A unique feature of the LHC is that large numbers of top quarks are produced with sufficient boosts that they  decay into single collimated jets on which jet shapes can be measured. In~\cite{Fleming:2007xt,Fleming:2007qr} it was shown using effective fields theories (SCET and bHQET)~\cite{Bauer:2001ct,Bauer:2000yr,Bauer:2001yt,Bauer:2002nz,Eichten:1989zv,Isgur:1989vq,Isgur:1990yhj,Grinstein:1990mj,Georgi:1990um,Manohar:2000dt} that factorization theorems can be derived for event shapes measured on boosted top quarks, enabling these observables to be expressed in terms of $m_t$ in a field theoretically well defined mass scheme~\cite{Hoang:2008yj,Hoang:2009yr,Hoang:2017suc,Hoang:2015vua,Bachu:2020nqn,Butenschoen:2016lpz,Hoang:2018zrp,ATLAS:2021urs}. Additionally, there has been substantial progress in parton shower algorithms capable of accurately simulating QCD radiation in fully exclusive top quark decays~\cite{Hoche:2017hno,Hoche:2017iem,Dulat:2018vuy,Dasgupta:2020fwr,Forshaw:2020wrq,Karlberg:2021kwr,Hamilton:2020rcu,Holguin:2020joq,Nagy:2020rmk,Brooks:2020upa,Hamilton:2021dyz,Bewick:2021nhc,Gellersen:2021eci,Forshaw:2021mtj,Frederix:2012ps,Hoeche:2014qda,Jezo:2016ujg,Frederix:2016rdc,Cormier:2018tog,Mazzitelli:2020jio}. In \Refcite{Hoang:2017kmk}, the groomed~\cite{Dasgupta:2013ihk,Larkoski:2014wba} jet mass was proposed as a $m_{t}$ sensitive observable, realizing the factorization based approach of~\cite{Fleming:2007xt,Fleming:2007qr}. For measurements, see~\cite{CMS:2017pcy,CMS:2019sic}. While jet grooming significantly improves the robustness of the observable, the complicated residual non-perturbative corrections~\cite{Hoang:2019ceu} continue to be limiting factors in achieving a precision competitive with direct measurements, thereby motivating the exploration of observables not reliant on grooming.

In recent years, there has been a program to rethink~\cite{Chen:2020vvp} jet substructure directly in terms of correlation functions, $\langle \cE(\vec n_1) \cdots \cE(\vec n_k) \rangle$, of the energy flow in a direction $\vec{n}$ ~\cite{Sveshnikov:1995vi,Tkachov:1995kk,Korchemsky:1999kt,Bauer:2008dt,Hofman:2008ar,Belitsky:2013xxa,Belitsky:2013bja,Kravchuk:2018htv}, $\mathcal{E}(\vec n)$, motivated by the original work in QCD~\cite{Basham:1978bw,Basham:1977iq,Basham:1979gh,Basham:1978zq,Konishi:1979cb,Tkachov:1994as,Tkachov:1999py,Grigoriev:2003yc,Korchemsky:1994is,Korchemsky:1997sy} and recent revival in conformal field theories (CFTs)~\cite{Hofman:2008ar,Belitsky:2013xxa,Belitsky:2013bja,Belitsky:2013ofa,Belitsky:2014zha,Korchemsky:2015ssa,Kravchuk:2018htv,Kologlu:2019bco,Kologlu:2019mfz,1822249,Korchemsky:2021okt,Korchemsky:2021htm}. These correlators have a number of unique and remarkable properties. Most importantly for phenomenological applications, correlators  are insensitive to soft radiation without the application of grooming. Additionally they can also be computed on tracks  \cite{Chen:2020vvp,Li:2021zcf,Jaarsma:2022kdd}, using the formalism of track functions~\cite{Chang:2013rca,Chang:2013iba}, allowing for higher angular resolution and suppressing pile-up. However, so far their application has been restricted to massless quark or gluon jets~\cite{Dixon:2019uzg,Chen:2019bpb,Chen:2020vvp,Chen:2020adz,Chen:2021gdk,Moult:2018jzp,Moult:2019vou,Gao:2019ojf,Ebert:2020sfi,Li:2020bub,Li:2021txc}.

In this \lett, we present the first steps towards a new paradigm for precision $m_{t}$ measurements based on the simple idea of exploiting the mass dependence of the characteristic opening \emph{angle} of the decay products of the boosted top, $\zeta \sim m_t^2/p_T^2$ (see \Fig{top_correlator}). 
The motivation for rephrasing the question in this manner is twofold. First, this angle can be accessed via low point correlators, which are field theoretically drastically more simple than a groomed substructure observable sensitive to $\zeta$. Second,  while the jet mass is sensitive to soft contamination and UE, the angle $\zeta$ is not, since it is primarily determined by the hard dynamics of the top decay. In the following, we will present a numerical proof-of-principles analysis illustrating that the three-point correlator in the vicinity of $\zeta \sim m_t^2/p_T^2$ provides a simple, but highly sensitive probe of $m_t$, free of the typical challenges of jet-shape based approaches. Our goal is to provide the motivation for future precision studies and the motivation to find solutions to outstanding theoretical problems in the study of low point correlators.

\section{The Three-Point Correlator}
\label{sec:3pt}
There has recently been significant progress in understanding the perturbative structure of correlation functions of energy flow operators. This includes the landmark calculation of the two-point correlator at next-to-leading order (NLO) in QCD~\cite{Dixon:2018qgp,Luo:2019nig} and NNLO in $\cN=4$ super Yang-Mills~\cite{Belitsky:2013ofa,Henn:2019gkr}, as well as the first calculation of a three-point correlator~\cite{Chen:2019bpb} at LO (also further analyzed in~\cite{Chen:2020adz,Chen:2021gdk,Karlberg:2021kwr}). The idea of using the three-point correlator to study the top quark is a natural one, and was considered early on in the jet substructure literature~\cite{Jankowiak:2011qa}. However, only due to this recent theoretical progress can we now make concrete steps towards a comprehensive program of using energy correlators as a precision tool for Standard Model measurements~\cite{Chen:2020vvp,EEC_OD}.

The three-point correlator (EEEC) with generic energy weights is defined, following the notation in~\cite{Chen:2019bpb}, as
\begin{align}
	G^{(n)}(\zeta_{12},\zeta_{23},\zeta_{31}) = \int \df \sigma \, \widehat{\cM}^{(n)}(\zeta_{12},\zeta_{23},\zeta_{31}) \, ,
\end{align}
with the measurement operator given by
\begin{align} \label{eq:measurement}
	&\widehat{\cM}^{(n)}(\zeta_{12},\zeta_{23},\zeta_{31}) =  \\
	& \sum_{i,j,k} \frac{E^{n}_{i}E^{n}_{j}E^{n}_{k}}{Q^{3n}} \delta\left(\zeta_{12} - \hat{\zeta}_{ij}\right)\delta\left(\zeta_{23} - \hat{\zeta}_{ik}\right)\delta\left(\zeta_{31}- \hat{\zeta}_{jk}\right)\,. \nn
\end{align}
Here $\hat{\zeta}_{ij}=(1-\cos(\theta_{ij}))/2$, with $\theta_{ij}$ the angle between particles $i$ and $j$, the sum runs over all triplets of particles in the jet, and $Q$ denotes the hard scale in the measurement. The EEEC is not an event-by-event observable, but rather is defined as an ensemble average.
\begin{figure}[t]
	\centering
	\includegraphics[width=0.45\textwidth]{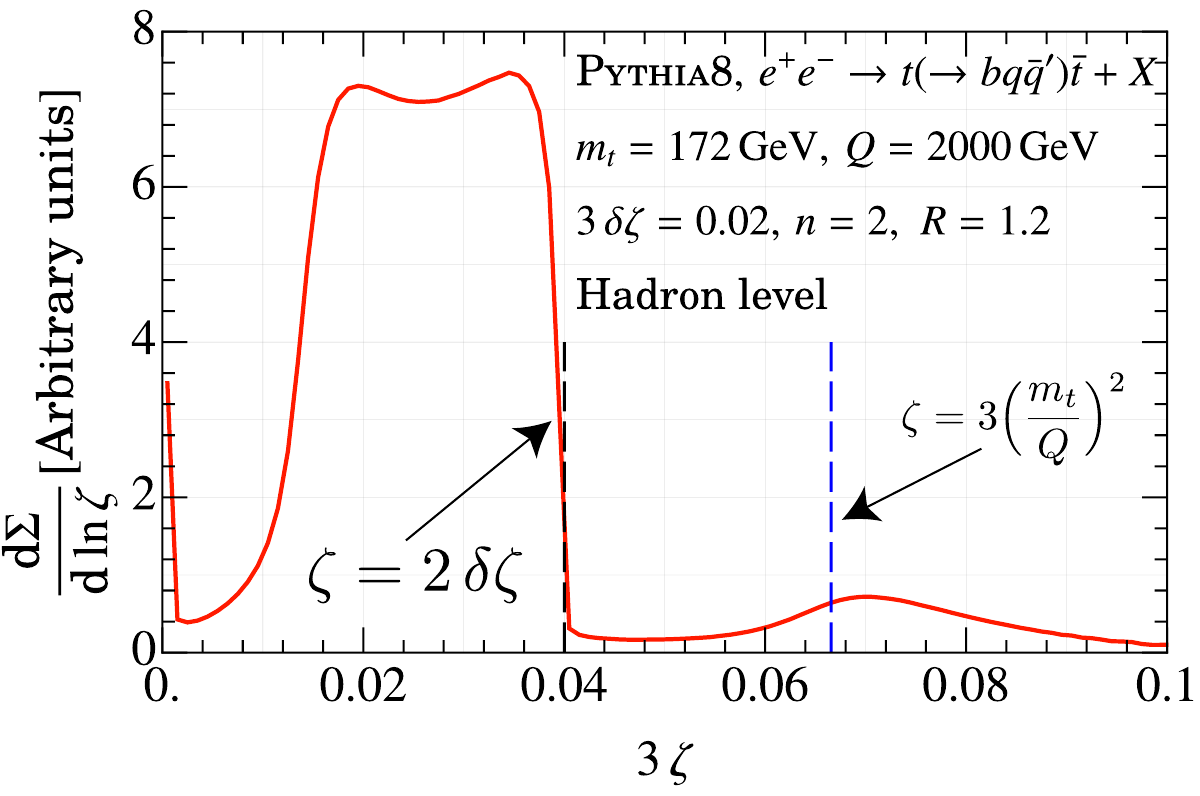}
	\caption{Features of the EEEC measurement in equilateral triangle configuration on the tops.}
	\label{fig:generalfeatures}
\end{figure}

We are interested in the limit $\zeta_{12},\zeta_{23},\zeta_{31}\ll 1$, such that all directions of energy flow lie within a single jet. In the case of a CFT (or massless QCD up to the running coupling), the EEEC simplifies  due to the rescaling symmetry along the light-like direction defining the jet. In this case, the EEEC can be written in terms of a scaling variable, $\zeta_{31}$ and exhibits a featureless power-law scaling governed by the twist-2 spin-4 anomalous dimension, $\gamma(4)$~\cite{Hofman:2008ar,Dixon:2019uzg,Korchemsky:2019nzm,Kologlu:2019mfz,Chen:2019bpb,Chen:2021gdk}. This behavior has been measured~\cite{EEC_OD} using publicly released CMS data~\cite{CMS:2008xjf,Komiske:2019jim}.

In contrast, $m_t$ explicitly breaks the rescaling symmetry of the collinear limit. Thus $m_t$ appears as a characteristic scale imprinted in the three-point correlator.  While the top quark has a three-body decay at leading order, higher-order corrections give rise to additional radiation, which is primarily collinear to the decay products leading to a growth in the distribution at angles $\hat{\zeta}_{ij} \ll m_t^2/p^{2}_{T}$. To extract $m_t$, we therefore focus on the correlator in a specific energy flow configuration sensitive to the hard decay kinematics. Here we study the simplest configuration, that of an equilateral triangle $\hat{\zeta}_{ij}=\zeta$ allowing for a small asymmetry ($\delta\zeta$). Thus the key object of our analysis is the $n^{\rm th}$ energy weighted cross section defined as
\begin{align}\label{eq:triangleEEEC}
	\frac{\df \Sigma (\delta \zeta)}{\df Q \df \zeta} =\! \int \df \zeta_{12} \df \zeta_{23} \df \zeta_{31} \! \int \df \sigma \widehat{\cM}^{(n)}_{\bigtriangleup}(\zeta_{12},\zeta_{23},\zeta_{31}, \zeta , \delta \zeta)\,,
\end{align} 
where the measurement operator $\widehat{\cM}^{(n)}_{\bigtriangleup}$ is
\begin{align} 
	&\widehat{\cM}^{(n)}_{\bigtriangleup}(\zeta_{12},\zeta_{23},\zeta_{31}, \zeta , \delta \zeta) = \widehat{\cM}^{(n)}(\zeta_{12},\zeta_{23},\zeta_{31}) \\
	& \times \delta (3\zeta - \zeta_{12} - \zeta_{23} - \zeta_{31})\!\!\! \prod_{l,m,n \in \{1,2,3\}}\!\! \Theta(\delta \zeta - |\zeta_{lm} - \zeta_{mn}| )\, .\nn
\end{align}
For $\delta \zeta \ll \zeta$,
\begin{align}
	\frac{\df \Sigma}{\df \zeta} \approx 4 (\delta \zeta)^{2} \, G^{(n)}(\zeta,\zeta,\zeta; m_{t}) \, ,
\end{align} 
where we have made the dependence on $m_t$ explicit. Three-body kinematics implies that the distribution is peaked at $ \zeta_{\text{peak}} \approx 3 m^2_t/Q^2$, exhibiting quadratic sensitivity to $m_{t}$.
At the LHC the peak is resilient to collinear radiation since $\ln \zeta_{\mathrm{peak}} < 1/\alpha_{\textrm{s}}$, makings its properties computable in fixed order perturbation theory at the hard scale. In the region $\zeta <2\delta \zeta $ the hard three-body kinematics is no longer identified, leading to a bulge in the distribution. In \Fig{generalfeatures} we show these features in the simplest case of $\ee \ra t + X$ simulated using \Pythiaxx parton shower, with the details of the simulation described below. We explain in \app{LO} through a leading-order analysis how these features arise and motivate the definition of our observable stated above.
Finally, we do not consider here the optimization of $\delta \zeta$ and leave it to future work.

\begin{figure}[t]
	\centering
	\includegraphics[width=0.45\textwidth]{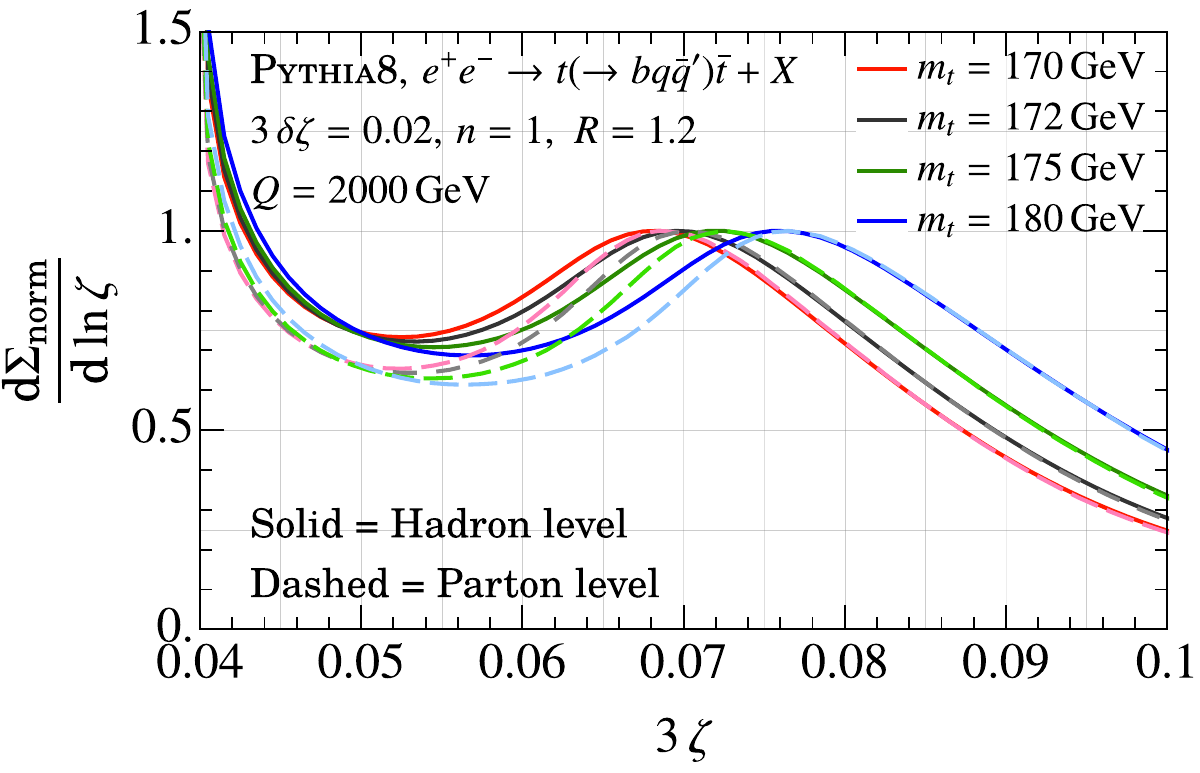}    
	\includegraphics[width=0.45\textwidth]{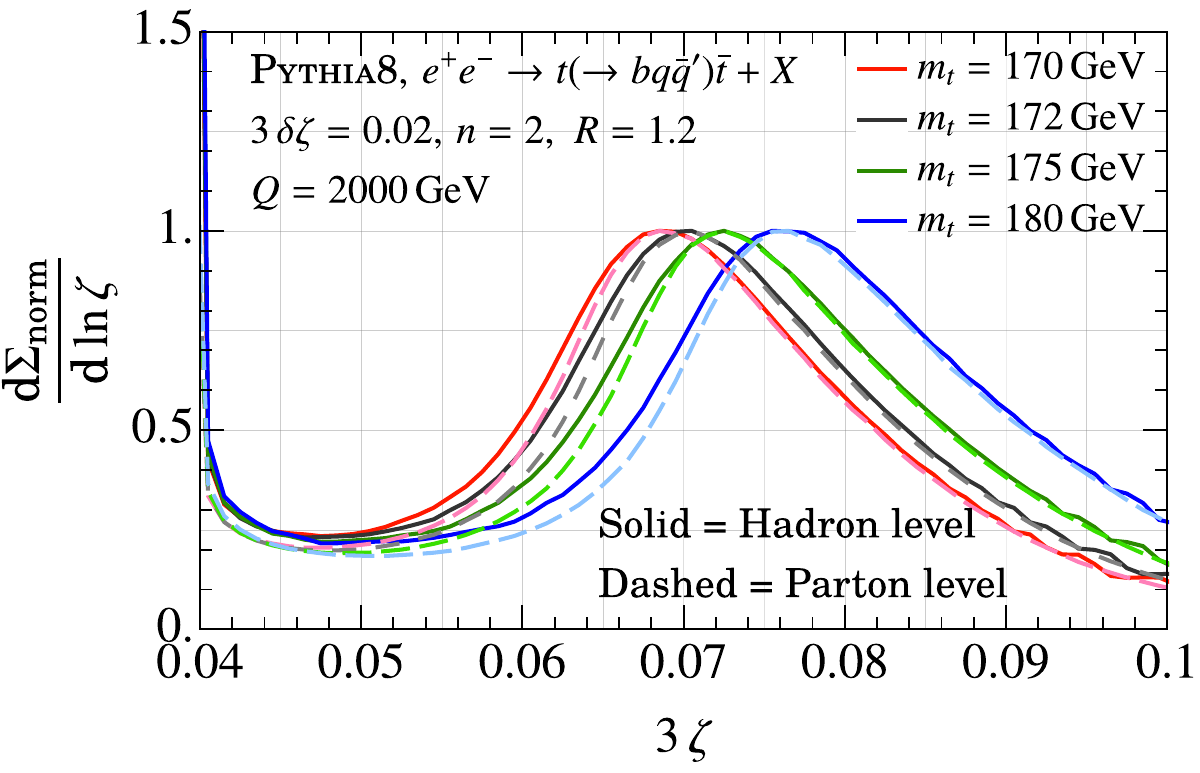} 
	\includegraphics[width=0.46\textwidth]{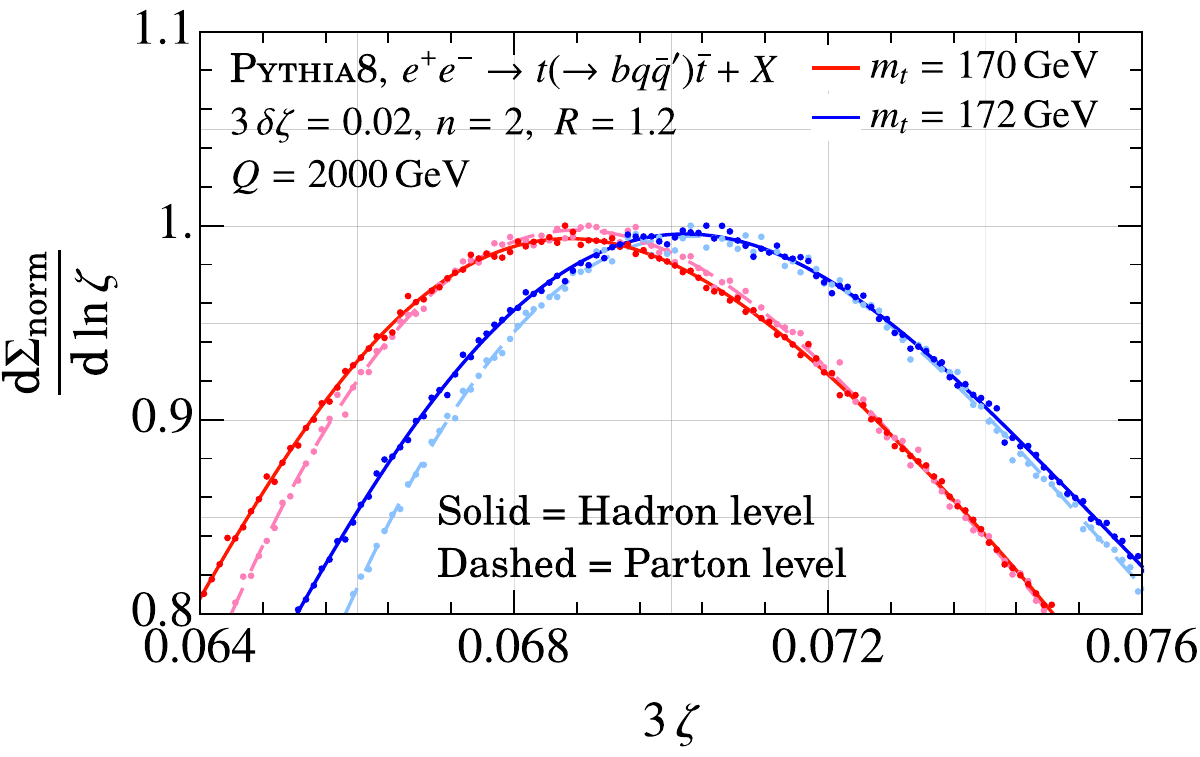}
	\caption{The $n=1,2$ three-point correlators on boosted tops in $e^+e^-$ showing a clear peak at $\zeta \sim 3m_t^2/Q^2$. All graphs are normalized to peak height. The bottom plot shows a zoomed-in version of the $n=2$ three-point correlator in $e^+e^-$ for $m_t=170,172$ GeV, at both hadron and parton level. The dashed and solid lines are a polynomial fit to Monte Carlo data points.
	}\label{fig:sensitivity}
\end{figure}

\section{Mass Sensitivity}
\label{sec:ee}

To illustrate the mass sensitivity of our observable, we consider the simplest case of $e^+e^-$ collisions simulated in \Pythiaxx at  a center of mass energy of $Q=2000$ GeV using the \Pythiaxx parton shower~\cite{Sjostrand:2014zea}. We reconstruct anti-$k_T$~\cite{Cacciari:2008gp} jets with $R=1.2$ using \Fastjet~\cite{Cacciari:2011ma}, and analyze them using the jet analysis software \texttt{JETlib}~\cite{jetlib}. Although jet clustering is not required in $e^+e^-$, this analysis strategy is chosen to achieve maximal similarity with the case of hadron colliders.
In \Fig{sensitivity} we show the distribution of the three-point correlator in the peak region, both with and without the effects of hadronization. Agreement of the peak position with the leading-order expectation is found, showing that the observed behavior is dictated by the hard decay of the top. In \Fig{sensitivity}, linear ($n = 1$) and quadratic ($n= 2$) energy weightings are used, see \eq{measurement}. The latter is not collinear safe, but the collinear IR-divergences can be absorbed into moments of the fragmentation functions or track functions~\cite{Chen:2020vvp,Li:2021zcf}. 

Non-perturbative effects in energy correlators are governed by an additive underlying power law~\cite{Korchemsky:1994is,Korchemsky:1997sy,Korchemsky:1999kt,Belitsky:2001ij}, which over the width of the peak has a minimal effect on the normalized distribution. This is confirmed by the small differences in peak position between parton and hadron level distributions. In \Fig{sensitivity} we also show a zoomed-in version for $n = 2$. 
Taking $m_t = 170 , 172$ GeV with $n=2$ as representative distributions, we find that the shift due to hadronization corresponds to a $\Delta m_t^{\rm Had.}\sim 250$ MeV shift in $m_{t}$. This is in contrast with the groomed jet mass case where hadronization causes peak shifts equivalent to $\Delta m_t^{\rm Had.} \sim 1$ GeV~\cite{Hoang:2017kmk}.

\section{Hadron Colliders}
\label{sec:pp}
We now extend our discussion to the more challenging case of proton-proton collisions. This study illustrates the difference between energy correlators and standard jet shape observables, and also emphasizes the irreducible difficulties of jet substructure at hadron colliders.

Implicit in the definition of energy correlators, $\langle \psi | \cE(\vec n_1) \cdots \cE(\vec n_k)|\psi \rangle$, is a characterization of the QCD final state $|\psi\rangle$. In the correlator literature, $|\psi\rangle$ is usually defined by a local operator of definite momentum acting on the QCD vacuum, $|\psi\rangle=\cO|0\rangle$, giving rise to a perfectly specified hard scale, $Q$. This is the case of $e^+e^-$ collisions. In hadronic final states at proton-proton collisions, the states on which we compute the energy correlators are necessarily defined through a measurement, e.g. by selecting anti-$k_T$ jets with a specific $p_{T,\text{jet}}$. Due to the insensitivity of the energy correlators to soft radiation, we will show that it is in fact the non-perturbative effects on the jet $p_T$ selection that are the only source of complications in a hadron collider environment. This represents a significant advantage of our approach, since it shifts the standard problem of characterizing non-perturbative corrections to infrared jet shape observables, to characterizing non-perturbative effects on a \emph{hard} scale. This enables us to propose a methodology for the precise extraction of $m_t$ in hadron collisions by independently measuring the universal non-perturbative effects on the $p_{T,{\rm jet}}$ spectrum. We now illustrate the key features of this approach.

The three-point correlator in hadron collisions is defined as
\begin{align} \label{eq:measurementLHC}
	&\widehat{\cM}_{(pp)}^{(n)}(\zeta_{12},\zeta_{23},\zeta_{31} ) = \sum_{i,j,k\,\in\, {\rm jet}} \frac{(p_{T,i})^{n}(p_{T,j})^{n}(p_{T,k})^{n}}{(p_{T,\text{jet}})^{3n}} \nn \\
	& \times \delta\left(\zeta_{12} - \hat{\zeta}_{ij}^{(pp)}\right)\delta\left(\zeta_{23} - \hat{\zeta}_{ik}^{(pp)}\right)\delta\left(\zeta_{31}- \hat{\zeta}_{jk}^{(pp)}\right)\,, 
\end{align}
where $\hat{\zeta}_{ij}^{(pp)} = \Delta R^{2}_{ij} = \Delta \eta_{ij}^2 + \Delta \phi_{ij}^2$, with $\eta,\phi$ the standard rapidity, azimuth coordinates. 
The peak of the EEEC distribution is determined by the hard kinematics and is found at $\zeta^{(pp)}_{\text{peak}}\approx 3m^2_t/p_{T, t}^2$, where $p_{T,t}$ is the hard top $p_{T}$, \textit{not} $p_{T, \mathrm{jet}}$.

\begin{figure}
	\includegraphics[width=0.45\textwidth]{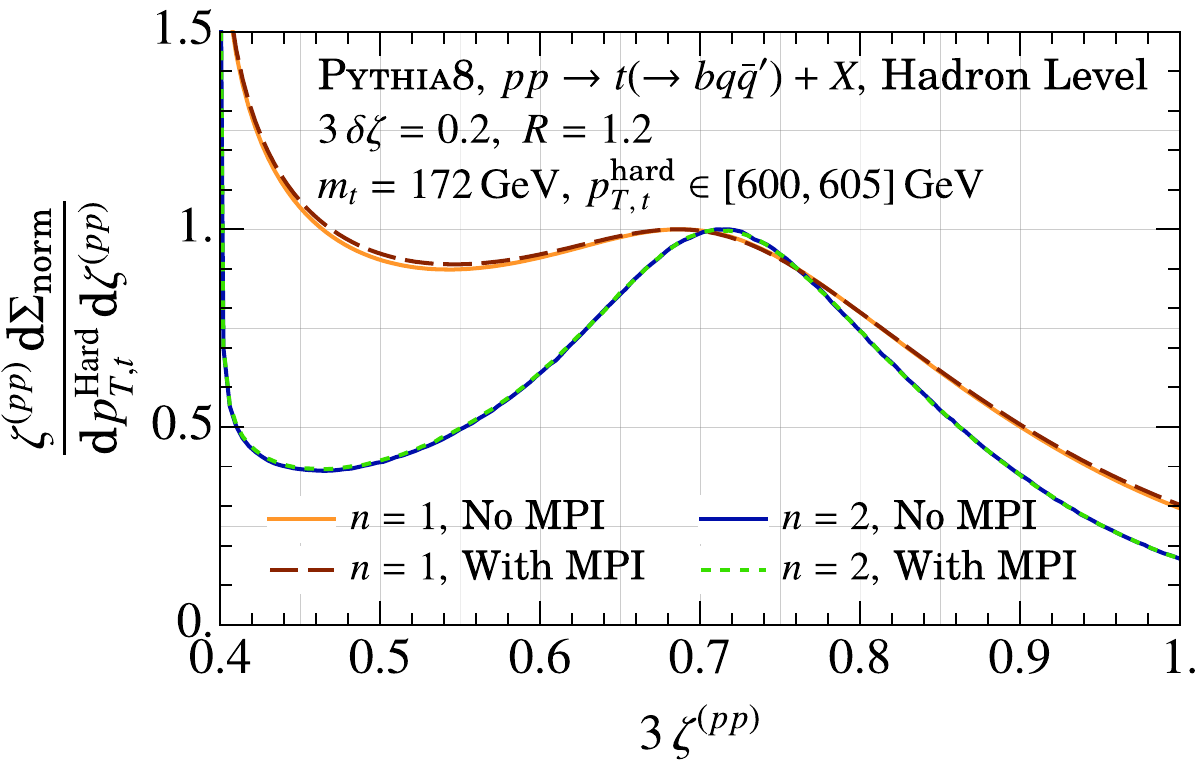}
	\caption{The $n=1,2$ three-point correlators on decaying top quarks with a fixed hard $p_T$, with and without MPI. Here a clear peak can be seen at $\zeta \approx 3m_t^2/p_{T,t}^2$.}
	\label{fig:pp_MPI_hard}
\end{figure}

To clearly illustrate the distinction between the infrared measurement of the EEEC and the hard measurement of the $p_{T,\text{jet}}$ spectrum, we present a two-step analysis using data generated in \Pythiaxx (which we independently verified with \Vinciaxx~\cite{vincia:2016}, see \Fig{pythiavsvincia} below). First, we generated hard top quark states with definite momentum (like in $e^+e^-$), but in the more complicated LHC environment including UE; shown in  \Fig{pp_MPI_hard}, where we see a clear peak that is \emph{completely} independent of the presence of MPI  (the \Pythiaxx model for UE). This illustrates that the correlators themselves, on a perfectly characterized top quark state, are insensitive to soft radiation without grooming.

We then performed a proof-of-principles analysis to illustrate that a characterization of non-perturbative corrections to the $p_{T,\text{jet}}$ spectrum allows us to extract $m_t$, with small uncertainties from non-perturbative physics. 
While we will later give a factorization formula for  the observable $\df \Sigma (\delta \zeta)/\df p_{T, \mathrm{jet}} \, \df \zeta$, for the present discussion it is useful to write it as 
\begin{align}
	\frac{\df \Sigma (\delta \zeta)}{\df p_{T, \mathrm{jet}} \, \df \zeta} = \frac{\df \Sigma (\delta \zeta)}{\df p_{T, t} \, \df \zeta} ~~ \frac{\df p_{T, t}}{\df p_{T, \mathrm{jet}}}\,.
\end{align}
This formula, combined with \Fig{pp_MPI_hard}, illustrates that the source of complications in the hadron-collider environment lies in the observable-independent function of \emph{hard scales} $\df p_{T, t}/\df p_{T, \mathrm{jet}}$, which receives both perturbative and non-perturbative contributions. 
To extract a value of $m_t$, we write the peak position as
\begin{align}\label{eq:pp_master}
	\zeta^{(pp)}_{\text{peak}}= \frac{3 F_\text{pert}(m_t, p_{T, \mathrm{jet}},\as,R)}{\left(p_{T, \mathrm{jet}} + \Delta_{\text{NP}}(R)+\Delta_\text{MPI}(R)\right)^2} \, .
\end{align}
Here $F_{\text{pert}}$ incorporates the effects of perturbative radiation. At leading order, $F_{\text{pert}}=m_t^2$. Corrections from hadronization and MPI are encoded through the shifts $\Delta_{\text{NP}}(R)$ and $\Delta_\text{MPI}(R)$. Crucially, in the factorization limit that we consider, these are not a property of the EEEC observable, but can instead be extracted directly from the non-perturbative corrections to the jet $p_T$ spectrum \cite{Dasgupta:2007wa}. This is a unique feature of our approach.

\begin{figure}
	\includegraphics[width=0.45\textwidth]{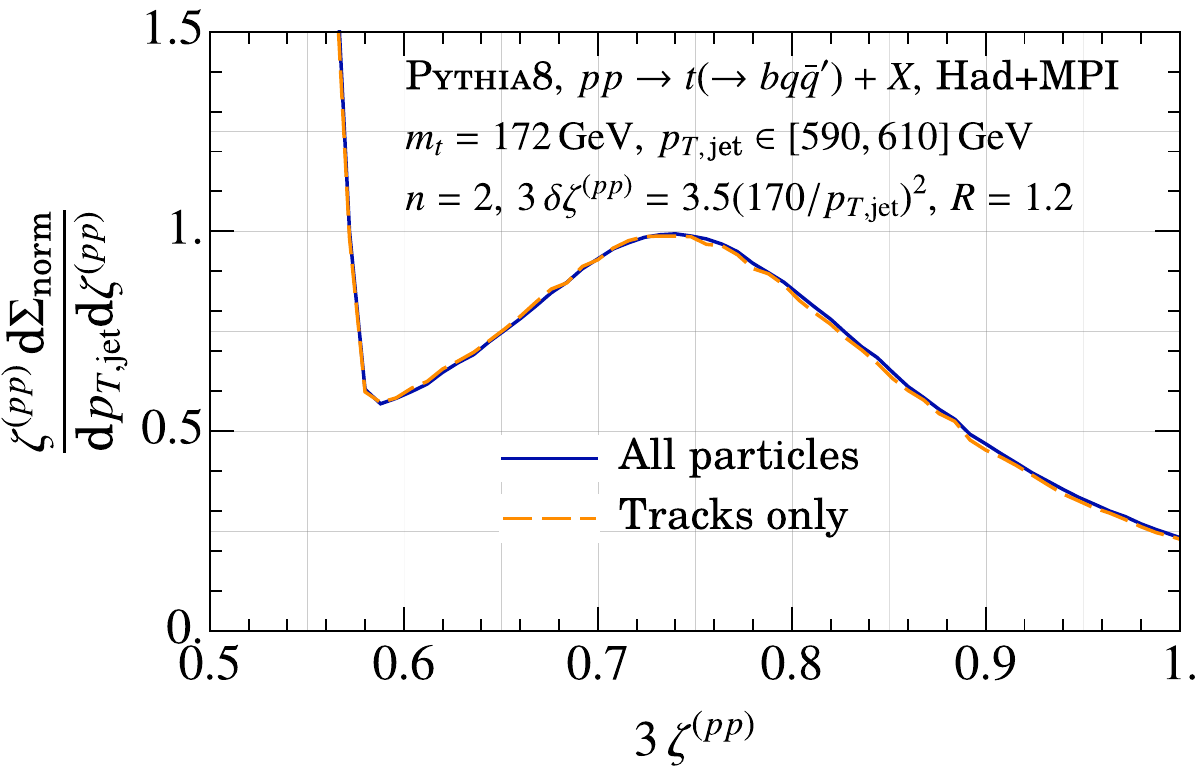}
	\caption{The three-point correlator on top jets in hadron collisions. A clear peak can be seen at $\zeta \approx 3m_t^2/p_{T,\text{jet}}^2$ which is insensitive to the usage of tracks.}
	\label{fig:pp_tracks}
\end{figure}

To illustrate the feasibility of this procedure, we used \Pythiaxx (including hadronization and MPI) to extract $\zeta^{(pp)}_{\text{peak}}$ as a function of $p_{T,\text{jet}}$, over an energy range within the  expected reach of the high luminosity LHC. As a proxy for a perturbative calculation, we used parton level data to extract $F_{\text{pert}}$. To the accuracy we are working, $F_{\text{pert}}$ is independent of the jet $p_T$, and can just be viewed as an effective top mass $\sqrt{F_{\text{pert}}}(m_{t})$. We also extract $\Delta_{\text{NP}}(R)+\Delta_\text{MPI}(R)$ independently from the $p_{T,\text{jet}}$ spectrum. Note that an error of $\pm \delta$ on $\Delta_{\text{NP}/\text{MPI}}$ in a given $p_{T,\mathrm{jet}}$ bin leads to an error on $\sqrt{F_{\text{pert}}}(m_{t})$ of $\pm \delta \sqrt{F_{\text{pert}}} / p_{T,\mathrm{jet}}$.

\begin{figure}
	\includegraphics[width=0.45\textwidth]{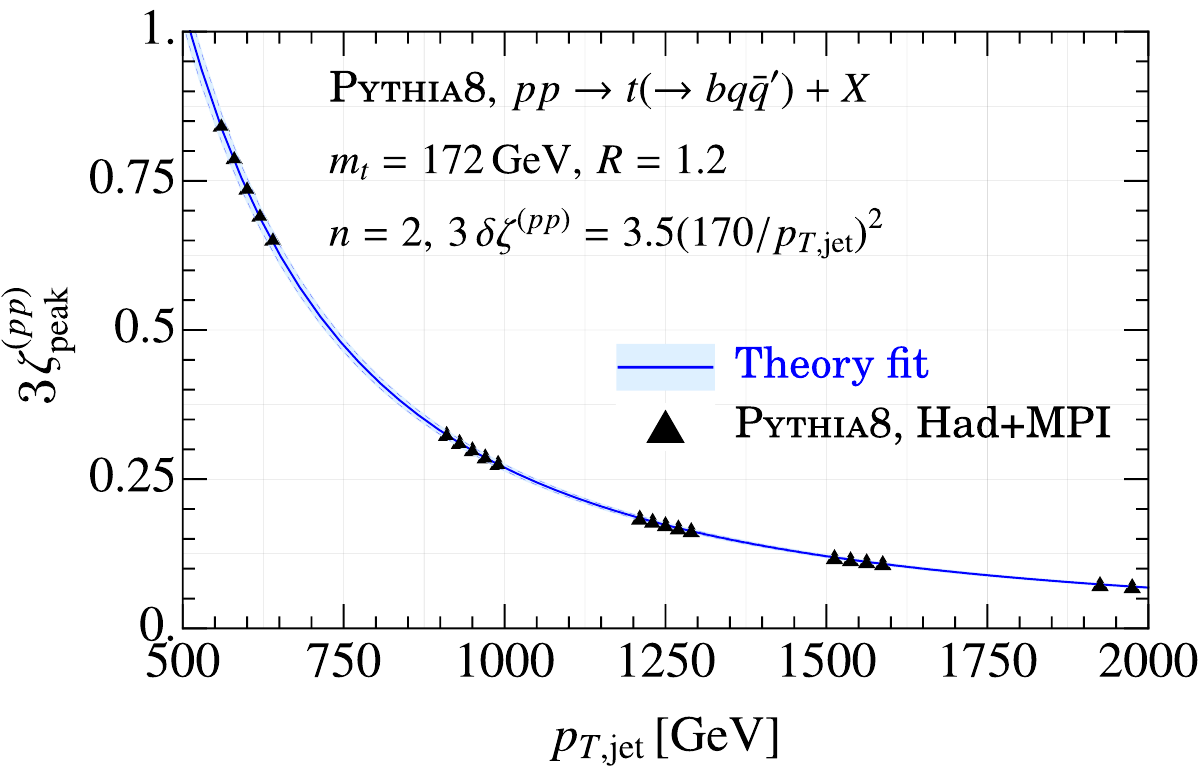}
	
	\caption{\label{fig:pp_MPI} Plot of the peak position as a function of $p_{T,\text{jet}}$, as used in our fitting procedure.}
\end{figure}

Using \eq{pp_master} we fit $\zeta^{(pp)}_{\text{peak}}$ as a function of $p_{T,{\rm jet}}$ for an effective value of $F_{\text{pert}}(m_t)$. An example of the distribution in the peak region is shown in \Fig{pp_tracks}, which also highlights the insensitivity of the peak position to the use of charged particles only (tracks). A fit to $\zeta^{(pp)}_{\text{peak}}$ for several $p_{T,{\rm jet}}$ bins is shown in \Fig{pp_MPI}. With a perfect characterization of the non-perturbative corrections to the EEEC observable, the value of $F_{\text{pert}}(m_t)$ extracted when hadronization and MPI are included should exactly match its extraction at parton level. This would lead to complete control over $m_{t}$. In \Tab{pp_table} we show the extracted value of $F_{\text{pert}}(m_t)$ from our parton level fit, and from our hadron+MPI level fit for two values of the \Pythiaxx $m_t$. The errors quoted are the statistical errors on the parton shower analysis. The Hadron+MPI fit is quoted with two errors: the first originates from the statistical error on the EEEC measurement, the second originates from the statistical error on the determination of $\Delta_{\text{NP}}(R)+\Delta_\text{MPI}(R)$ from the $p_{T,\text{jet}}$ spectrum. A more detailed discussion of this procedure is provided in \app{analysis}. Thus we find promising evidence that theoretical control of $m_{t}$, with conservative errors $\lesssim 1$GeV, is possible with an EEEC-based measurement. Our analysis also emphasizes the importance of understanding non-perturbative corrections to the jet $p_T$ spectrum. Theory errors are contingent upon currently unavailable NLO computations, discussed in the following section, and so are not provided. However, we expect observable dependent NLO theory errors on $m_{t}$ to be better than those in other inclusive measurements wherein in the dominant theory errors are from PDFs$+\alpha_{\textrm{s}}$ \cite{CMS:2014rml,CMS:2017xrt} and which mostly affect the normalisation of the observable. By contrast the EEEC is also inclusive but the extracted $m_{t}$ is only sensitive to the observable's shape.

The goal of this \lett has been to introduce our novel approach to top mass measurements, illustrating its theoretical feasibility and advantages. Our promising results motivate developing a deeper theoretical understanding of the three-point correlator of boosted tops in the hadron collider environment. Nevertheless, there remain many areas in which our methodology could be improved to achieve greater statistical power and bring it closer to experimental reality. These include the optimization of $\delta \zeta$, the binning of $p_{T,\text{jet}}$ and $\zeta^{(pp)}$, and including other shapes on the EEEC correlator. Regardless, our analysis does demonstrate the observable's potential for a precision $m_{t}$ extraction when measured on a sufficiently large sample of boosted tops. We are optimistic that such a sample will be accessible at the HL-LHC where it is forecast that $\sim 10^7$ boosted top events with $p_T>500\,$GeV will be measured~\cite{Azzi:2019yne}. 

\begin{table}
	\begin{tabular}{||c|c|c||}
		\hline
		\Pythiaxx $m_{t}$ & Parton $\sqrt{F_{\text{pert}}}$ & Hadron + MPI $\sqrt{F_{\text{pert}}}$ \\ \hline
		$172$ GeV & $172.6\pm 0.3$ GeV &$172.3 \pm 0.2 \pm 0.4$ GeV\\
		$173$ GeV & $173.5 \pm 0.3$ GeV  & $173.6\pm 0.2 \pm 0.4$ GeV\\
		$175$ GeV & $175.5 \pm 0.4$ GeV  & $175.1\pm 0.3 \pm 0.4$ GeV\\ \hline
		$173-172$ & $0.9 \pm 0.4$ GeV & $1.3\pm 0.6 $ GeV\\ 
		$175-172$ & $2.9 \pm 0.5$ GeV & $2.8\pm 0.6 $ GeV\\ \hline
	\end{tabular}
	\caption{\label{tab:pp_table}The effective parameter $F_{\text{pert}}(m_t)$ extracted at parton level, and hadron+MPI level. The consistency of the two simulations provides a measure of our uncertainty due to uncontrolled non-perturbative corrections. Statistical errors are shown. }
\end{table}

\section{Factorization Theorem}
\label{sec:fact}
Combining factorization for massless energy correlators~\cite{Dixon:2019uzg} with the bHQET treatment of the top quark near its mass shell~\cite{Fleming:2007xt,Fleming:2007qr,Hoang:2017kmk,Bachu:2020nqn} allows us to separate the dynamics at the scale of the hard production, the jet radius $R$, the angle $\zeta$, and the top width $\Gamma_t$.  While factorization is generically violated for hadronic jet shapes (see~\cite{Forshaw:2021fxs}), our framework is based on the rigorous factorization for single particle massive fragmentation~\cite{Collins:1981ta,Bodwin:1984hc,Collins:1985ue,Collins:1988ig,Collins:1989gx,Collins:2011zzd,Nayak:2005rt,Mitov:2012gt}. Assuming $\zeta \ll R$, we perform a matching at  the perturbative scale of the jet radius, using the fragmenting jet formalism~\cite{Procura:2009vm,Kang:2016mcy,Kang:2016ehg}, which captures the jet algorithm dependence. The final jet function describing the collinear dynamics at the scale of $\zeta$ is therefore free of any jet algorithm dependence.
Correspondingly, we expect to obtain the following factorized expression 
\begin{align}\label{eq:factorization}
	&\frac{\df\Sigma}{\df p_{T,\text{jet}} \df \eta\, \df \zeta}=\,  f_i \otimes f_j \otimes H_{i,j\to t}\Big(z_J;p_{T,t} = \frac{p_{T,{\rm jet}}}{z_J},\eta \Big)\nn \\
	&\quad\otimes J_{t\to t}(z_J,z_{h};R)\otimes J_{\text{EEEC}}^{\rm{[tracks]}}(n,z_{h}, \zeta;m_t;\Gamma_t)\,, 
\end{align}
for the energy-weighted cross section differential in $p_{T,\text{jet}}$, rapidity $\eta$, and $\zeta$.  This can be used to compute $F_{\text{pert}}(m_{t})$ in a systematically improvable fashion. Obvious dependencies, such as on factorization scales, have been suppressed for compactness. Here $f_i$ are parton distribution functions, and $H_{i,j\to t}$ is the hard function for inclusive massive fragmentation~\cite{Mele:1990yq,Mele:1990cw}, which is known for LHC processes at NNLO~\cite{Czakon:2021ohs}. $J_{t\to t}$ is the fragmenting jet function, which is known at NLO for anti-$k_T$ jets~\cite{Kang:2016mcy,Kang:2016ehg}, but can be extended to NNLO using the approach of~\cite{Liu:2021xzi}. The convolutions over $f_{i,j}$ $H_{i,j\to t}$ and $J_{t\to t}$ alone determine the $p_{T,\text{jet}}$ spectrum, independent of the EEEC measurement. Finally, $J_{\text{EEEC}}$, is the energy correlator jet function, which can be computed in a well defined short-distance top mass scheme (such as the MSR mass~\cite{Hoang:2008yj,Hoang:2017suc,Hoang:2017btd}), and can include information from track or fragmentation functions. Around the top peak, $J_{\text{EEEC}}$ is almost entirely determined by perturbative physics and is currently known at LO. The NLO determination of $J_{\text{EEEC}}$ is an outstanding theoretical problem and is very involved, thus beyond the scope of this \lett, though a road map towards its completion has recently become available \cite{Dixon:2018qgp,Luo:2019nig,Chen:2019bpb}. In the region of on-shell top, $J_{\text{EEEC}}$ can be matched onto a jet function defined in bHQET~\cite{Fleming:2007xt,Fleming:2007qr,Hoang:2017kmk,Hoang:2008yj,Hoang:2009yr}. The functions in the factorization formula above exhibit standard DGLAP~\cite{Dokshitzer:1977sg,Gribov:1972ri,Altarelli:1977zs}  evolution in the momentum fractions $z_J$  and $z_{h}=p_{T, \text{hadron}} / p_{T,\text{jet}}$, and the $\otimes$ denote standard fragmentation convolutions. A more detailed study of the structure of the factorization will be provided in a future publication.

\section{Conclusions}
\label{sec:conclusion}

We have proposed a new paradigm for jet-substructure based measurements of the top mass at the LHC in a rigorous field theoretic setup. Instead of using standard jet shape observables, we have analyzed the three-point correlator of energy flow operators, and have illustrated a number of its remarkable features.  Our results support the possibility of achieving complete theoretical control over an observable with top mass sensitivity competitive with direct measurements whilst avoiding the ambiguities associated with the usage of MC event generators. 

\begin{acknowledgments}
We are particularly grateful to G. Salam for insightful questions that led us to (hopefully) significantly improve our presentation. We also thank B. Nachman, M. Schwartz, I. Stewart and W. Waalewijn for feedback on the manuscript. We thank H. Chen, P. Komiske,  K. Lee,  Y. Li, F. Ringer, J. Thaler, A. Venkata, X.Y. Zhang and H.X. Zhu for many useful discussions and collaborations on related topics that significantly influenced the philosophy of this work. A.P. is grateful to M. Vos, M. LeBlanc, J. Roloff and J. Aparisi-Pozo for many helpful discussions about subtleties of the top mass extraction at the LHC. This work is supported in part by the GLUODYNAMICS project funded by the ``P2IO LabEx (ANR-10-LABX-0038)'' in the framework ``Investissements d’Avenir'' (ANR-11-IDEX-0003-01) managed by the Agence Nationale de la Recherche (ANR), France. I.M. is supported by startup funds from Yale University. A.P. is a member of the Lancaster-Manchester-Sheffield Consortium for Fundamental Physics, which is supported by the UK Science and Technology Facilities Council (STFC) under grant number ST/T001038/1.
\end{acknowledgments}

\appendix

\section{Leading-Order Analysis}
\label{app:LO}
\begin{figure}[t]
	\includegraphics[width=0.45\textwidth]{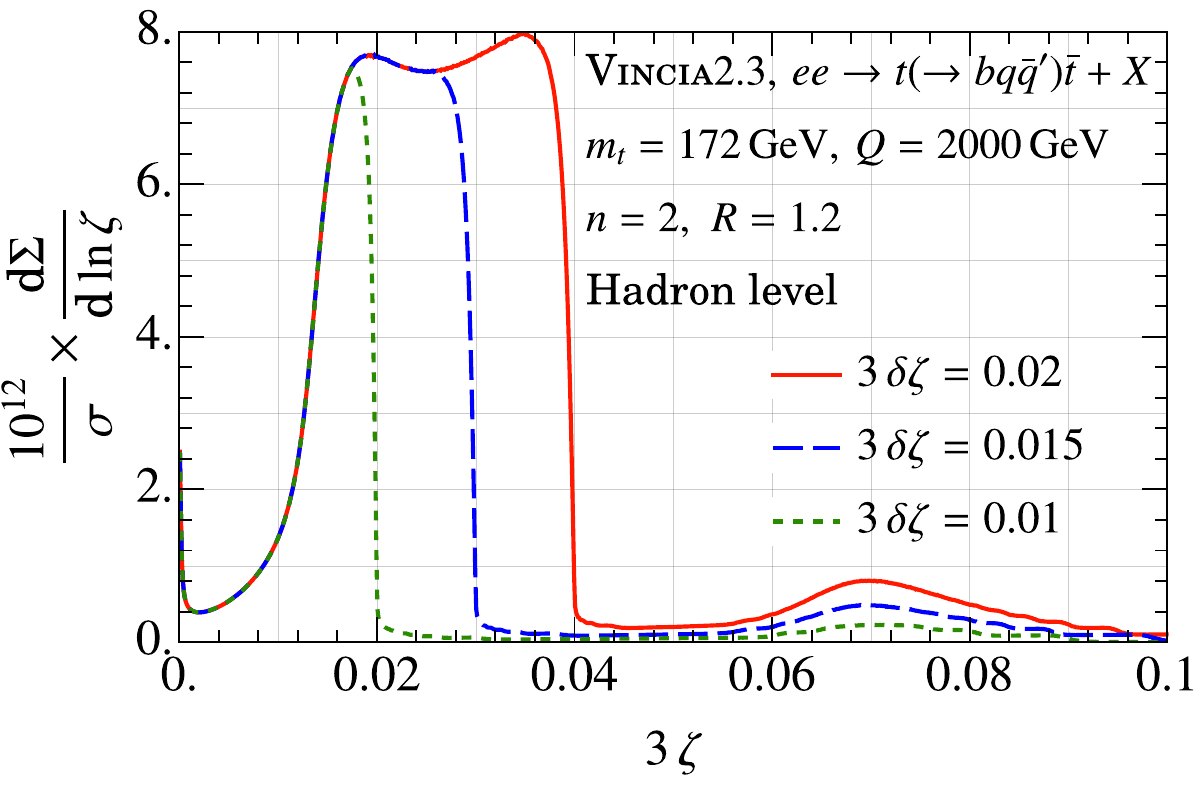} 
	\includegraphics[width=0.45\textwidth]{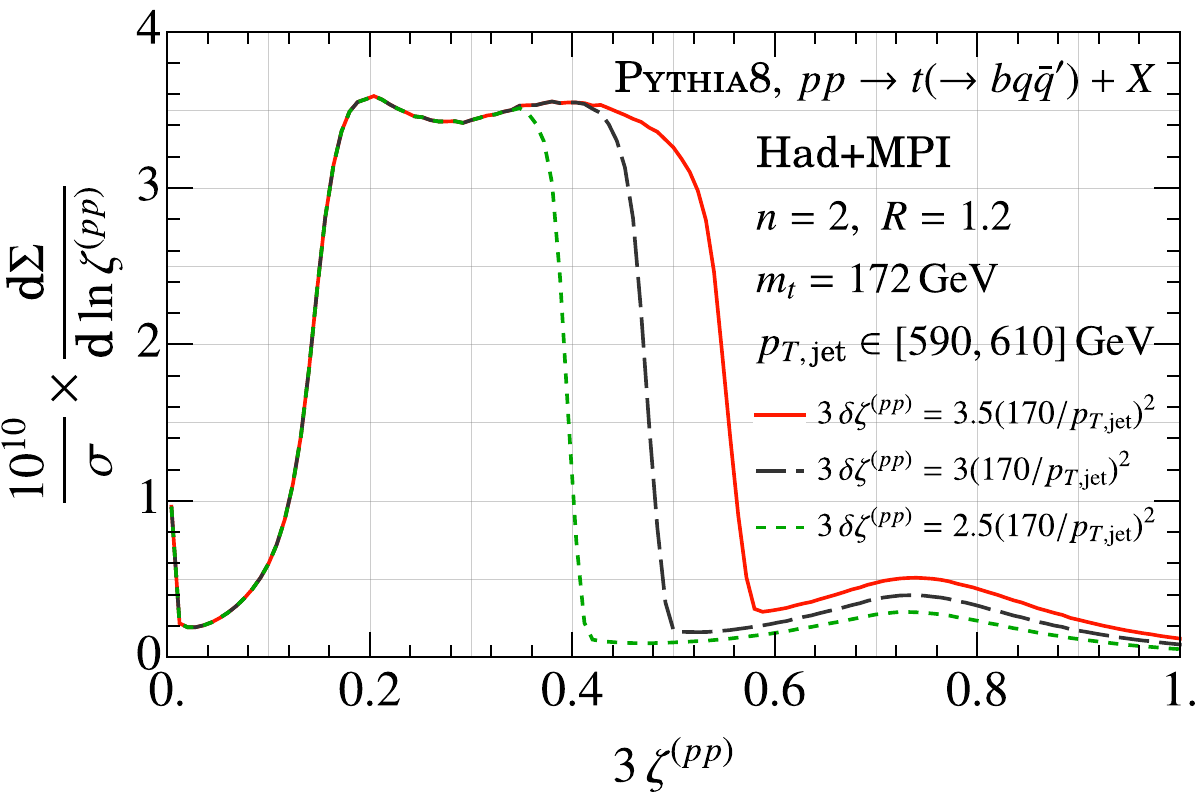} 
	\caption{The effect of applying different $\delta \zeta$ cuts to ensure an equilateral configuration for $\ee\rightarrow t + X$ and $pp \ra t + X$ processes. The $\delta\zeta$ cuts isolate the peak, which is governed by the hard decay of the top, from the ``bulge'' contribution.}
	\label{fig:asym}
\end{figure}

Here we perform a leading-order analysis of the observable which suffices to explain the general features of the spectrum in \Fig{generalfeatures}.
For concreteness, we will define the kinematics assuming a $e^{+}e^{-} \rightarrow t(\rightarrow b q \bar{q}') + X$ process where we take the $b, q, \bar{q}'$ partons to be massless. No further complications, beyond the need for more ink, are introduced by using the longitudinally invariant kinematics needed for measurements at the LHC.
At leading order, we can factorize the Born cross-section $\df \sigma^{(0)}/\sigma^{(0)}$ into the dimensionless three-body phase space for the top's decay products, $\df \Phi_{3}$, and the dimensionless weighted squared matrix element, $\sigma_{t}|M(t\rightarrow b W \rightarrow b q \bar{q}')|^{2}/\sigma^{(0)}$ where $\sigma_{t}$ is the cross section to produce a top quark. As $|M(t\rightarrow b W \rightarrow b q \bar{q}')|^{2} \sim \cO(1)$, we can approximate the differential EEEC distribution in \eq{triangleEEEC} as 
\begin{align}
	\frac{1}{\sigma^{(0)}}\frac{\df \Sigma^{(0)}}{\df Q \df \zeta_{12} \df \zeta_{23} \df \zeta_{31}} \approx \int \df \Phi_{3}\: \widehat{\cM}^{(n)}(\zeta_{12},\zeta_{23},\zeta_{31}) \, , \label{eq:approx3body}
\end{align} 
reducing the problem of understanding the observable of interest to studying three-body kinematics.

Before directly working with \eq{approx3body}, let us develop some intuition for the three-body kinematics. Consider the decay of a top quark in its rest frame, with $\tilde{p}_{t} = \tilde{p}_{b}  + \tilde{p}_{q} + \tilde{p}_{\bar{q}'}$.
Here we are using $\tilde{p}_i$ as a rest-frame momentum and $p_i$ as a lab-frame momentum. In the top rest frame, the angular parameters on which the EEEC depends are given by
\begin{align}
	\tilde{\zeta}_{12} = \frac{\tilde{p}_{b} \cdot  \tilde{p}_{q}}{2\tilde{E}_{b}\tilde{E}_{q}}, \quad \tilde{\zeta}_{31} = \frac{\tilde{p}_{b} \cdot  \tilde{p}_{\bar{q}'}}{2\tilde{E}_{b}\tilde{E}_{\bar{q}'}}, \quad \tilde{\zeta}_{23} = \frac{\tilde{p}_{q} \cdot  \tilde{p}_{\bar{q}'}}{2\tilde{E}_{q}\tilde{E}_{\bar{q}'}} \, .
\end{align}
Momentum conservation requires that $\tilde{\zeta}_{12}+\tilde{\zeta}_{23}+\tilde{\zeta}_{31} \in [2,2.25]$. Let the lab frame top momentum be $p_{t} = (E_{t},\vec{p}_{t})$. In the boost between the lab and rest frame, $\cosh \beta = E_{t}/ m_{t} \sim \zeta^{-1/2}$. To first order in $m_{t}/ E_{t} \ll 1$, we also have $\sinh{\beta} \approx \cosh \beta$. Hence the sum of lab frame EEEC parameters is
\begin{align}
	&\zeta_{12} + \zeta_{23} + \zeta_{31} \\
	&\quad \nn \approx \left(\frac{m_{t}}{E_{t}}\right)^{2}\left(\tilde{x}_{tb}\tilde{x}_{tq} \tilde{\zeta}_{12}+\tilde{x}_{tb}\tilde{x}_{t\bar{q}'}\tilde{\zeta}_{31}+\tilde{x}_{tq}\tilde{x}_{t\bar{q}'}\tilde{\zeta}_{23} \right),
\end{align}
where
\begin{align}
	\tilde{x}_{ti}=(1 + \cos \tilde{\theta}_{ti}) \, ,
\end{align}
with $ \tilde{\theta}_{ti}$ denoting the angle between parton $i$ and the boost axis in the top's rest frame. The function $$g \equiv \tilde{x}_{tb}\tilde{x}_{tq} \tilde{\zeta}_{12}+\tilde{x}_{tb}\tilde{x}_{t\bar{q}'}\tilde{\zeta}_{31}+\tilde{x}_{tq}\tilde{x}_{t\bar{q}'}\tilde{\zeta}_{23}$$ is also kinematically bounded so that $g \in [0,3]$. Upon averaging over the possible boost axes one finds that  $\left< g \right> \in [1,2.25]$. Thus, returning to \eq{approx3body}, we expect the partially integrated EEEC distribution
\begin{align}
	\frac{\df \Sigma}{\df Q \df \zeta} =& \int  \df \zeta_{12} \df \zeta_{23} \df \zeta_{31}\\
	&\quad \nn  \times \frac{\df \Sigma^{(0)}}{\df Q  \df \zeta_{12} \df \zeta_{23} \df \zeta_{31}} \delta (3\zeta - \zeta_{12} - \zeta_{23} - \zeta_{31}) \, ,
\end{align} 
to be peaked around $\zeta \approx \left< g \right>  m_{t}^{2}/(3E_{t}^{2}) \approx 2m_{t}^{2}/(3E_{t}^{2})$. However, this peak will have a large width (of the order of $3 m_{t}^{2}/(4 E_{t}^{2})$), whose origin can be understood by interpreting the parameters $\tilde{x}_{ti} \in [0,2]$ as three sources of (correlated) random noise in the shape of the flow of energy which `smears' the EEEC distribution. We can largely remove the noise by constraining the shape of the energy flow on the celestial sphere. This is most simply done by requiring that $\sqrt{\zeta_{ij}}$ approximately form the sides of an equilateral triangle ($\sqrt{\zeta_{ij}} \approx \sqrt{\zeta_{ik}}$). Consequently,
\begin{align}
	\tilde{x}_{tb}\tilde{x}_{tq} \tilde{\zeta}_{12} \approx \tilde{x}_{tb}\tilde{x}_{t\bar{q}'}\tilde{\zeta}_{31} \approx \tilde{x}_{tq}\tilde{x}_{t\bar{q}'}\tilde{\zeta}_{23} \, ,
\end{align}
removing two of the noisy degrees of freedom from the distribution. Upon including this constraint, we find that $\left< g \right> \approx 2.1$ with a small variance. This motivates us to introduce an EEEC distribution on equally spaced triplets of partons and allow for small asymmetries around this configuration governed by the parameter $\delta \zeta$:
\begin{align}\label{eq:triangleEEEC2}
	\frac{\df \Sigma (\delta \zeta)}{\df Q \df \zeta} = \int \df \zeta_{12} \df \zeta_{23} \df \zeta_{31} \int \df \sigma \widehat{\cM}^{(n)}_{\bigtriangleup}(\zeta_{12},\zeta_{23},\zeta_{31}, \zeta , \delta \zeta) \, ,
\end{align} 
where the operator $\widehat{\cM}^{(n)}_{\bigtriangleup}$ in the collinear limit is
\begin{align} 
	&\widehat{\cM}^{(n)}_{\bigtriangleup}(\zeta_{12},\zeta_{23},\zeta_{31}, \zeta , \delta \zeta) = \\
	&\quad \nn \sum_{i,j,k} \frac{E^{n}_{i}E^{n}_{j}E^{n}_{k}}{Q^{3n}} \delta\Big(\zeta_{12} - \frac{\theta^{2}_{ij}}{4}\Big)\delta\Big(\zeta_{31} - \frac{\theta^{2}_{ik}}{4}\Big)\delta\Big(\zeta_{23} - \frac{\theta^{2}_{jk}}{4}\Big)  \\
	& \quad \times \delta (3\zeta - \zeta_{12} - \zeta_{23} - \zeta_{31}) \hspace{-15pt}\prod_{l,m,n \in \{1,2,3\}}\hspace{-5pt} \Theta(\delta \zeta - |\zeta_{lm} - \zeta_{mn}| ).\nn
\end{align}
As previously explained, three-body kinematics determines that this distribution is peaked at $\zeta_{\mathrm{peak}} \approx 3 m_{t}^{2}/(4E_{t}^{2}) \sim m_{t}^{2}/Q^{2}$. Furthermore, at the LHC the peak should be resilient to collinear radiation since $\ln \zeta_{\mathrm{peak}} < 1/\as$.
\begin{figure}
	\centering
	\includegraphics[width=0.45\textwidth]{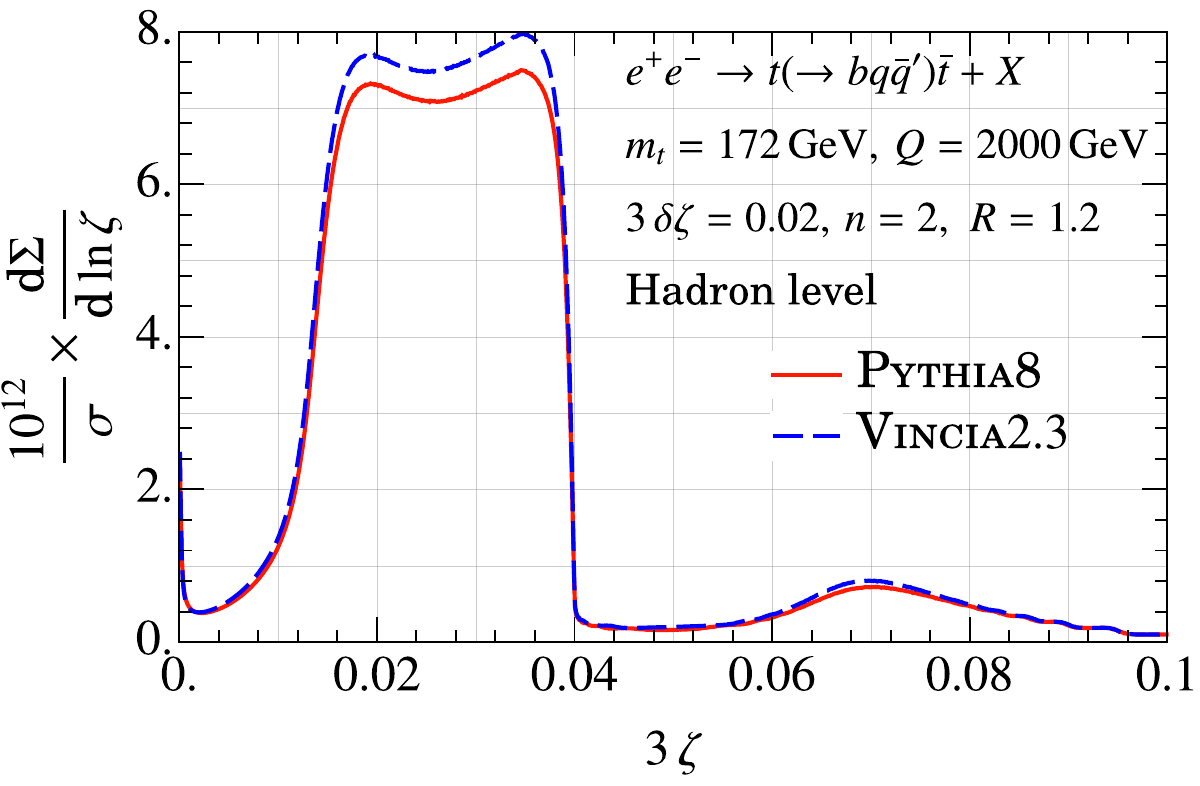}
	\caption{Comparison of \Pythiaxx and \Vinciaxx parton showers result. The differences in the peak positions are less than 300 MeV, and hence compatible with the uncertainties of our analysis.}
	\label{fig:pythiavsvincia}
\end{figure}

We can now complete our leading-order discussion by computing the Born contribution to \eq{triangleEEEC2}. Expanding for $\delta \zeta \ll \zeta$, we obtain 
\begin{align}
	\frac{\df \Sigma^{0} (\delta \zeta)}{\df Q \df \zeta } &\propto  (\delta \zeta)^{2} \int^{1}_{0} \df z_{1} \df z_{2} \df z_{3} \left(\frac{z_{1}z_{2}z_{3}}{8}  \right)^{n} \\
	&\quad \times \delta\left(\frac{m^{2}_{t}}{4E_{t}^{2}} - z_{1}z_{2}\zeta-z_{1}z_{3}\zeta-z_{2}z_{3}\zeta\right) \nn \\
	& \quad \times \delta (1-z_{1}-z_{2}-z_{3}) \, |M(t\rightarrow b W \rightarrow b q \bar{q}')|^{2} \, ,\nn
\end{align} 
where $z_{1} = E_{b}/E_{t}$ and $z_{2} = E_{\bar{q}'}/E_{t}$. The delta function causes the distribution to be sharply peaked at $\zeta = 3m^{2}_{t}/(4E_{t}^{2})$. This matches the intuition we have developed from considering pure kinematics. 

Looking at \eq{triangleEEEC} to all orders in $\as$, up to power corrections in $\delta \zeta$,
\begin{align}
	\frac{\df \Sigma(\delta \zeta)}{\df Q\df \zeta } =  4 (\delta \zeta)^{2} \, G^{(n)}(\zeta,\zeta,\zeta)\bigg(1 +  \cO\Big(\frac{\delta \zeta}{\zeta}\Big)\bigg)\, ,
\end{align} 
where the latter to leading order in $\delta \zeta \ll \zeta$ can be written as,
\begin{align}
	4 (\delta \zeta)^{2} \, G^{(n)}(\zeta,\zeta,\zeta) = \frac{\df \Sigma^{0} (\delta \zeta)}{\df Q \df \zeta } + \cO(\as)	\,	,
\end{align}
whilst in the region where $2 \delta \zeta > \zeta$
\begin{align}
	\frac{\df \Sigma(\delta \zeta)}{\df Q \df \zeta } \approx \int \df \zeta_{12} \df \zeta_{23} \df \zeta_{31} \, G^{(n)}(\zeta_{12},\zeta_{23},\zeta_{31}) \,	.
\end{align} 
\Fig{asym} demonstrates that this dependence on $\delta \zeta$ is born out in simulation for $\ee$ and $pp$ collisions.
To conclude our discussion of $\delta \zeta$, the limiting cases discussed above motivate that an optimal choice of this parameter will be a function of $Q$ that strikes a balance between statistics and constraining the three-body kinematics ($\delta \zeta_{\mathrm{optimal}} \approx \kappa\, \zeta_{\mathrm{peak}}/2$ for $\kappa \lesssim 1$). A more sophisticated analysis may also sum over several shapes of energy flow on the celestial sphere to increase statistics --- perhaps allowing for smaller values of $\delta \zeta_{\mathrm{optimal}}$.

\begin{figure}[t]
	\centering
	\includegraphics[width=0.45\textwidth]{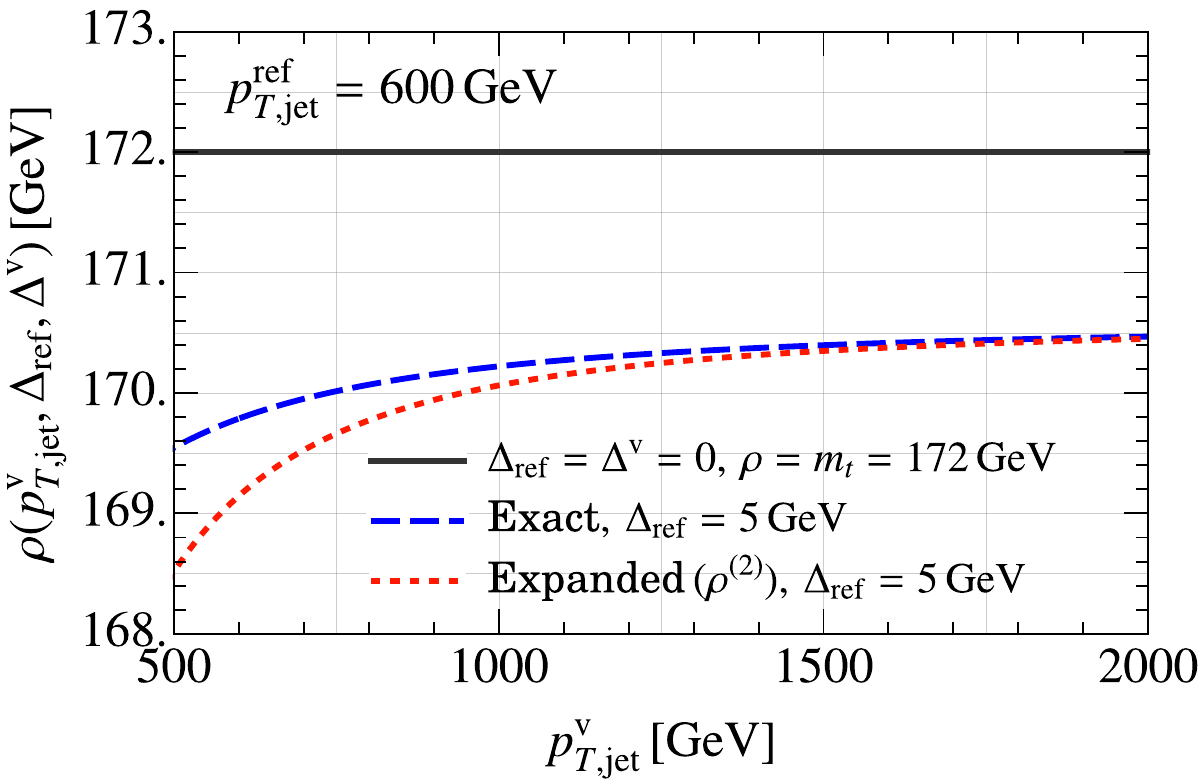}
	\caption{The function $\rho$ defined \eq{y} for $\Delta^{\text{ref}}=0$ GeV, as well as the exact and expanded to second order ($\rho^{(2)}$) for $\Delta^{\text{ref}}=5$ GeV.}
	\label{fig:rhoexp}
\end{figure}
Finally, in \Fig{pythiavsvincia} we show the top peak in the 3-point correlator for $n = 2$ in $\ee \ra t + X$ simulations in \Vinciaxx. We find the peak position almost in line with that of \Pythiaxx, justifying our earlier assumption that  the features of the observable are largely determined by the fixed-order expansion in $\as$.

\section{Details of the EEEC Analysis at Hadron Colliders}
\label{app:analysis}

Here we describe the details of the proof-of-principles peak position analysis outlined in \secn{pp}.
The longitudinally boost invariant measurement operator for the EEEC observable is
\begin{align}
	&\widehat{\cM}_{(pp)}^{(n)}(\zeta_{12},\zeta_{23},\zeta_{31} ) =\hspace{-5pt} \sum_{i,j,k\,\in\, {\rm jet}}\hspace{-5pt} \frac{(p_{T,i})^{n}(p_{T,j})^{n}(p_{T,k})^{n}}{(p_{T,{\rm jet}})^{3n}}\\
	&\quad \times \nn  \delta\left(\zeta_{12} - \hat{\zeta}_{ij}^{(pp)}\right)\delta\left(\zeta_{23} - \hat{\zeta}_{ik}^{(pp)}\right)\delta\left(\zeta_{31}- \hat{\zeta}_{jk}^{(pp)}\right)\,, 
\end{align}
where $\hat{\zeta}_{ij}^{(pp)} = \Delta R^{2}_{ij} = \Delta \eta_{ij}^2 + \Delta \phi_{ij}^2$. As before, the peak of the $\widehat{\cM}^{(n)}_{\bigtriangleup}$ EEEC distribution is determined by the top quark hard kinematics and is found at $\zeta^{(pp)}_{\text{peak}}\approx 3m^2_t/p_{T, t}^2$, where $p_{T,t}$ is the hard top $p_{T}$, \textit{not} $p_{T, \mathrm{jet}}$. Consequently, the basic properties of the $\df \Sigma (\delta \zeta)/\df p_{T, t} \, \df \zeta$ distribution are completely insensitive to non-perturbative physics. In \secs{ee}{pp} we demonstrated this insensitivity by parton shower simulation wherein we showed evidence that the top decay peak is nearly entirely independent of hadronization and UE. Consequently, in the limit that $p_{T, t}/(\Delta_{\text{NP}}+\Delta_{\text{MPI}})\rightarrow \infty $, the top decay peak position is exactly independent of non-perturbative effects. However, since $p_{T,t}$ is not directly accessible, the observable we consider is 
\begin{align}
	\frac{\df \Sigma (\delta \zeta)}{\df p_{T, \mathrm{jet}} \, \df \zeta} = \frac{\df \Sigma (\delta \zeta)}{\df p_{T, t} \, \df \zeta} ~~ \frac{\df p_{T, t}}{\df p_{T, \mathrm{jet}}}\,,
\end{align}
where $p_{T, \mathrm{jet}}$ is the $p_T$ of an identified anti-$k_T$ top-jet. The top peak position in the distribution $ \df \Sigma (\delta \zeta)/\df p_{T, \mathrm{jet}} \, \df \zeta$ will be shifted by hadronization and UE due to shifts in the jet $p_{T}$ distribution. This shift can be measured independently from our observable and will be universal to all measurements of energy correlators on top quarks at the LHC.

\begin{figure}
	\centering
	\includegraphics[width=0.45\textwidth]{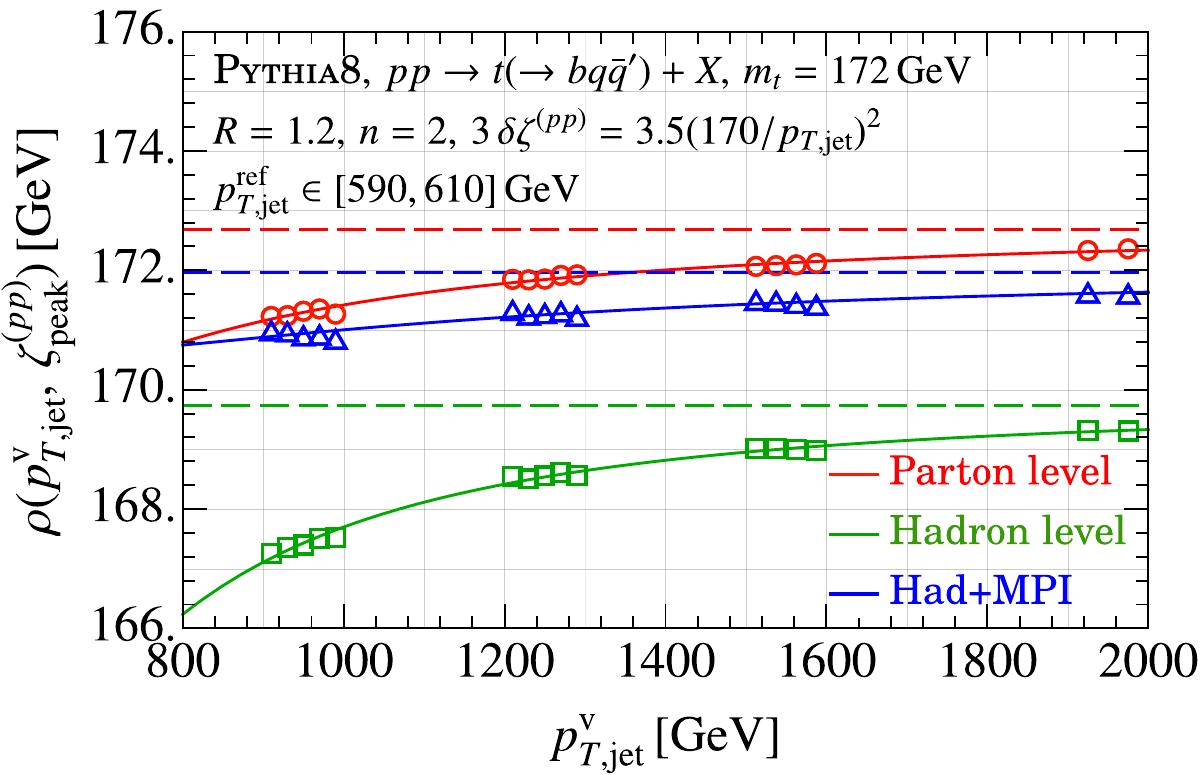}
	\caption{
		An example of the best fit for $\rho$'s asymptote ($\rho_{\text{asy}}$) using the fit  function in \eq{rhofit}. The data being fitted is produced using \Pythiaxx with $m_{t}=172$ and with $p^{\text{ref}}_{T,\text{jet}}$ binned in the range $p^{\text{ref}}_{T,\text{jet}}\in [590,610]$GeV . The dashed lines are the best fit for the asymptotes, $\rho =\rho_{\text{asy}}$.}
	\label{fig:asyFit}
\end{figure}

We can parameterize the all-orders peak position in $\df \Sigma (\delta \zeta)/\df p_{T, \mathrm{jet}} \, \df \zeta$ as
\begin{align}\label{eq:allorderspeak}
	\zeta^{(pp)}_{\text{peak}}&= 3(1 + \cO(\as))\frac{m^2_t}{f(p_{T, \mathrm{jet}},m_{t},\as,\LQCD)^2} \\
	&\nn\equiv 3(1 + \cO(\as))\frac{m^2_t}{\left(p_{T, \mathrm{jet}}+\Delta(p_{T, \mathrm{jet}},m_{t},\as,\LQCD)\right)^2}	\,	.
\end{align}
Mainly, $\Delta$ receives three additive contributions from perturbative radiation, hadronization, and from UE/MPI:
\begin{align}
	\Delta = \Delta_{\text{pert}} + \Delta_{\text{NP}} + \Delta_\text{MPI}.
\end{align}
Some simple manipulations can be made so as to minimize the sensitivity to $\Delta$ in an extracted value of $m_t$. We define the following function of measurable and perturbatively calculable quantities,
\begin{align}\label{eq:y}
	&\rho^2(\zeta^{(pp)\text{v}}_{\text{peak}},p^{\text{v}}_{T, \mathrm{jet}}) = \left(\zeta^{(pp)\text{ref}}_{\text{peak}} - \zeta^{(pp)\text{v}}_{\text{peak}}\right)\\
	&\qquad\nn \times \left(\frac{3(1 + \cO(\as))}{(p^{\text{v}}_{T, \mathrm{jet}})^2} - \frac{3(1 + \cO(\as))}{(p^{\text{ref}}_{T, \mathrm{jet}})^2} \right)^{-1},
\end{align}
where $\zeta^{(pp)\text{ref}}_{\text{peak}}$ is the peak position in a fixed reference $p_T$ bin, $p^{\text{ref}}_{T, \mathrm{jet}}$, and $\zeta^{(pp)\text{v}}_{\text{peak}}$ is the peak position for a variable $p_{T,{\rm jet}}$ value, $p^{\text{v}}_{T, \mathrm{jet}}$, larger than the reference value (we require $p^{\text{v}}_{T, \mathrm{jet}}> p^{\text{ref}}_{T, \mathrm{jet}}$ to avoid divergences). $\rho^2$ is defined so that, in the limit $p^{\text{v}}_{T, \mathrm{jet}},p^{\text{ref}}_{T, \mathrm{jet}} \rightarrow \infty$, we have $\rho^2 \rightarrow m^{2}_{t}$. In the analysis below we set $3(1 + {\cal O}(\alpha_s))\mapsto 3$ so that, in the limit $p^{\text{v}}_{T, \mathrm{jet}},p^{\text{ref}}_{T, \mathrm{jet}} \rightarrow \infty$, we find $\rho^2 \rightarrow F_{\text{pert}}$ as defined in \eq{pp_master}. Now let us make a further definition, 
\begin{align}
	& \Delta^{\text{v}}(p^{\text{v}}_{T, \mathrm{jet}} - p^{\text{ref}}_{T, \mathrm{jet}},m_{t},\as,\LQCD) \\
	&\qquad \equiv \Delta(p^{\text{v}}_{T, \mathrm{jet}},m_{t},\as, \LQCD)  - \Delta^{\text{ref}} \, ,\nn
\end{align}
where
\begin{align}\label{eq:DeltaRefDef}
	\Delta^{\text{ref}}&\equiv \Delta(p^{\text{ref}}_{T, \mathrm{jet}},m_{t},\as,\LQCD)	\,	.
\end{align}
We can substitute \eq{allorderspeak} into \eq{y} to find $\rho(p^{\text{v}}_{T, \mathrm{jet}}, \Delta^{\text{ref}},\Delta^{\text{v}})$, which is plotted in \Fig{asyFit}, left.

$\rho$ has an asymptote as $p^{\text{v}}_{T, \mathrm{jet}} \rightarrow \infty$ around which we perform a series expansion:
\begin{widetext}
\begin{center}
	\begin{figure}[t]
		\centering
		\includegraphics[width=0.45\textwidth]{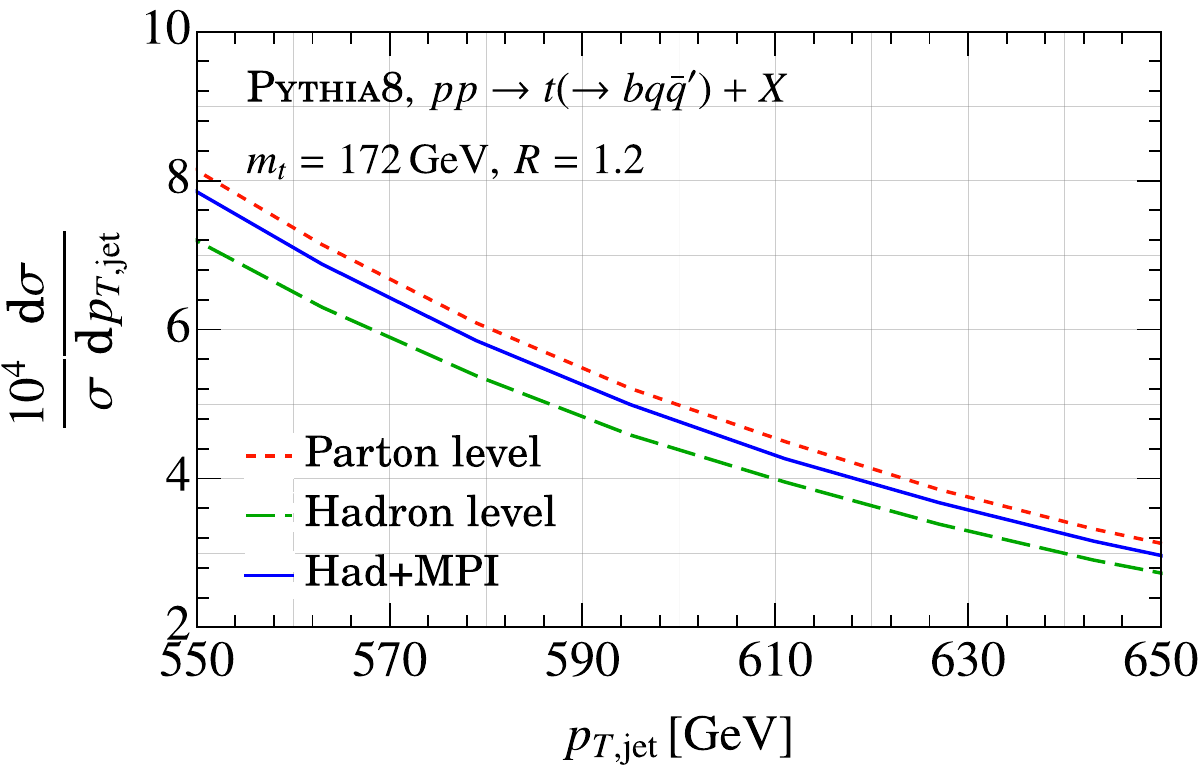}
		\includegraphics[width=0.45\textwidth]{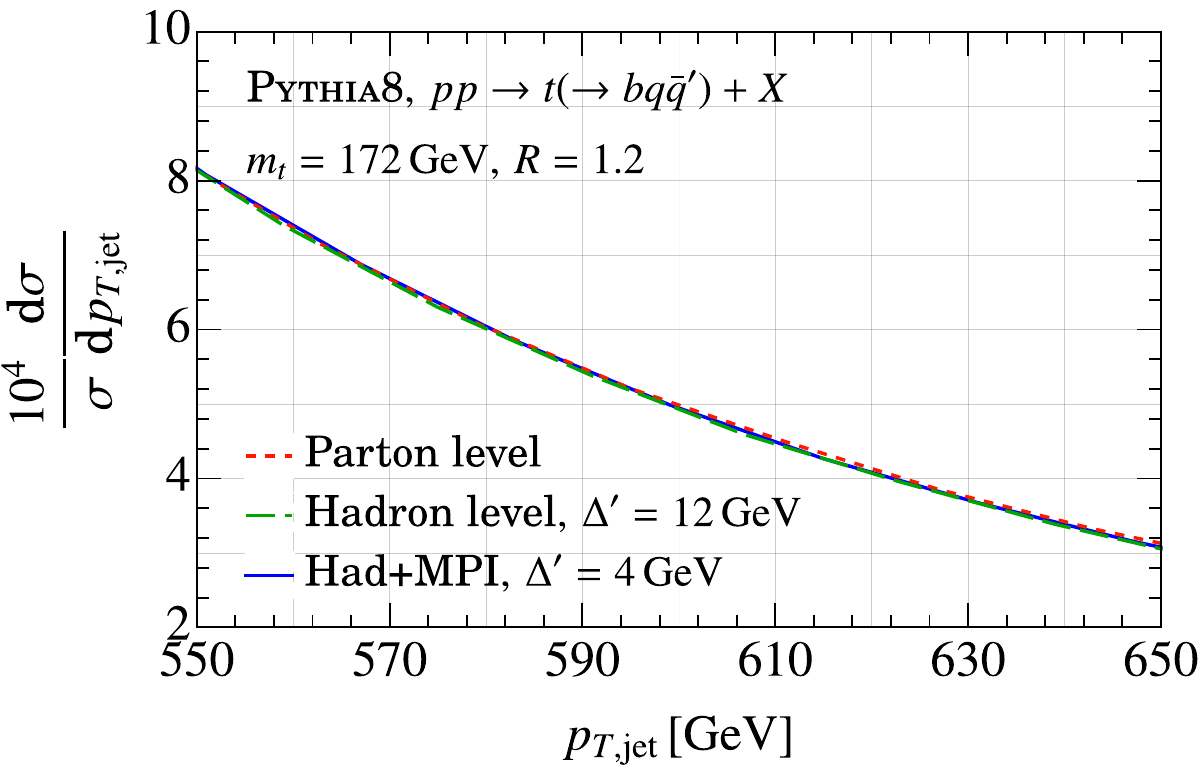}
		\caption{On the left, the $p_{T,\text{jet}}$ spectrum at parton level, hadron level, and including MPI. A precise characterization of the states on which the energy correlator is computed requires an understanding of the non-perturbative shifts between these distributions. On the right, the hadron level and hadron+MPI curves shifted by constant values so that the three curves overlap. In both figures $p_{T,\text{jet}} \in [550,650]$GeV as this is the range in which we chose $p^{\text{ref}}_{T,\text{jet}}$ in our analysis.}
		\label{fig:pt_shifts}
	\end{figure}
\end{center}
	\begin{align}
		\rho(p^{\text{v}}_{T, \mathrm{jet}}, \Delta^{\text{ref}},\Delta^{\text{v}}) =& \sqrt{F_{\text{pert}}} \frac{p^{\text{ref}}_{T, \mathrm{jet}}}{p^{\text{ref}}_{T, \mathrm{jet}}  + \Delta^{\text{ref}}}\Bigg(1- \frac{2p^{\text{ref}}_{T, \mathrm{jet}}\Delta^{\text{ref}} + (\Delta^{\text{ref}})^{2}}{2(p^{\text{v}}_{T, \mathrm{jet}})^{2}} + \frac{\left(p^{\text{ref}}_{T, \mathrm{jet}} + \Delta^{\text{ref}} \right)^{2}\left(\Delta^{\text{ref}} + \Delta^{\text{v}} \right)}{8(p^{\text{v}}_{T, \mathrm{jet}})^{3}} + \cO\Big(\frac{1}{(p^{\text{v}}_{T, \mathrm{jet}})^{4}}\Big) \Bigg)\, . \label{eq:rhoexpansion}
	\end{align}
\end{widetext}
Thus a fit of the asymptote of $\rho$, and its first non-zero correction, can be used to extract $F_{\text{pert}}$ and $\Delta^{\text{ref}}$. All dependence on $\Delta^{\rm v}$ enters in the higher order terms. However, in the limit that $p^{\text{ref}}_{T, \mathrm{jet}} \rightarrow \infty$, $\rho^2 \rightarrow F_{\text{pert}}$ and so while the fit for $F_{\text{pert}}$ will become exact, the error on a fit for $\Delta^{\text{ref}}$ will diverge. In practice it will be necessary to perform the EEEC measurement with boosted tops in order to get a well-defined peak. Consequently, fits for $\Delta^{\text{ref}}$ will suffer from parametrically large errors (as can be seen in the large deviation between the exact and expanded curves at low $p^{\text{ref}}_{T, \mathrm{jet}}$ in \Fig{rhoexp}). However, as previously stated, $\Delta^{\text{ref}}$ can be extracted from an independent measurement of the top-jet $p_T$ distribution,
\begin{align}
	\frac{\df \sigma_{pp\rightarrow t(\rightarrow b q \bar{q}') + X}}{\df p^{\text{ref}}_{T, \mathrm{jet}}}=D(m_{t},p_{T, \mathrm{jet}},\as,\LQCD)	\,	.
\end{align}
One can parameterize the non-pertubative effects in $D$ in the same way as we did in $\zeta^{(pp)}_{\text{peak}}$ to give 
\begin{align}
&D(m_{t},p_{T, \mathrm{jet}},\as,\LQCD) \\
&\quad\nn= D^{\rm pert.}(m_{t},g(p_{T, \mathrm{jet}},m_{t},\as,\LQCD),\as)	\,	,
\end{align}
where $D^{\rm pert.}(m_{t},p_{T, \mathrm{jet}},\as)$ is the all-orders perturbative top-jet $p_T$ distribution, and $g(p_{T,{\rm jet}}, \ldots)$ captures all the non-perturbative modifications to $p_{T,{\rm jet}}$. As before, we parameterize the modifications via introducing a shift function $\Delta'$ defined as
\begin{align}
	&\Delta'(p_{T, \mathrm{jet}},m_{t},\as,\LQCD) \\
	&\qquad \equiv g(p_{T, \mathrm{jet}},m_{t},\as,\LQCD)\nn
	-p_{T, \mathrm{jet}} \, , 
\end{align}
where
\begin{align}\label{eq:DeltaPrimeDef}
		\Delta'^{\text{ref}} &\equiv 
	g(p^{\text{ref}}_{T, \mathrm{jet}},m_{t},\as,\LQCD)
	-p^{\text{ref}}_{T, \mathrm{jet}} \,.
\end{align}
It is required for consistency with the factorization in \eq{factorization} that, up to corrections which are suppressed by powers of $m_t/p^{\text{ref}}_{T, \mathrm{jet}}$ and $\LQCD/p^{\text{ref}}_{T, \mathrm{jet}}$, $\Delta'^{\text{ref}} \approx (\Delta^{\text{ref}} - \Delta^{\text{ref}}_{\text{pert}})$ where $\Delta^{\text{ref}}_{\text{pert}}$ is the perturbative contribution to $\Delta^{\text{ref}}$. At the level of accuracy to which we are working, $\Delta^{\text{ref}}_{\text{pert}}$ can be absorbed into $F_{\text{pert}}$ justifying why we dropped it above.

\begin{figure}[t]
	\centering
	\includegraphics[width=0.45\textwidth]{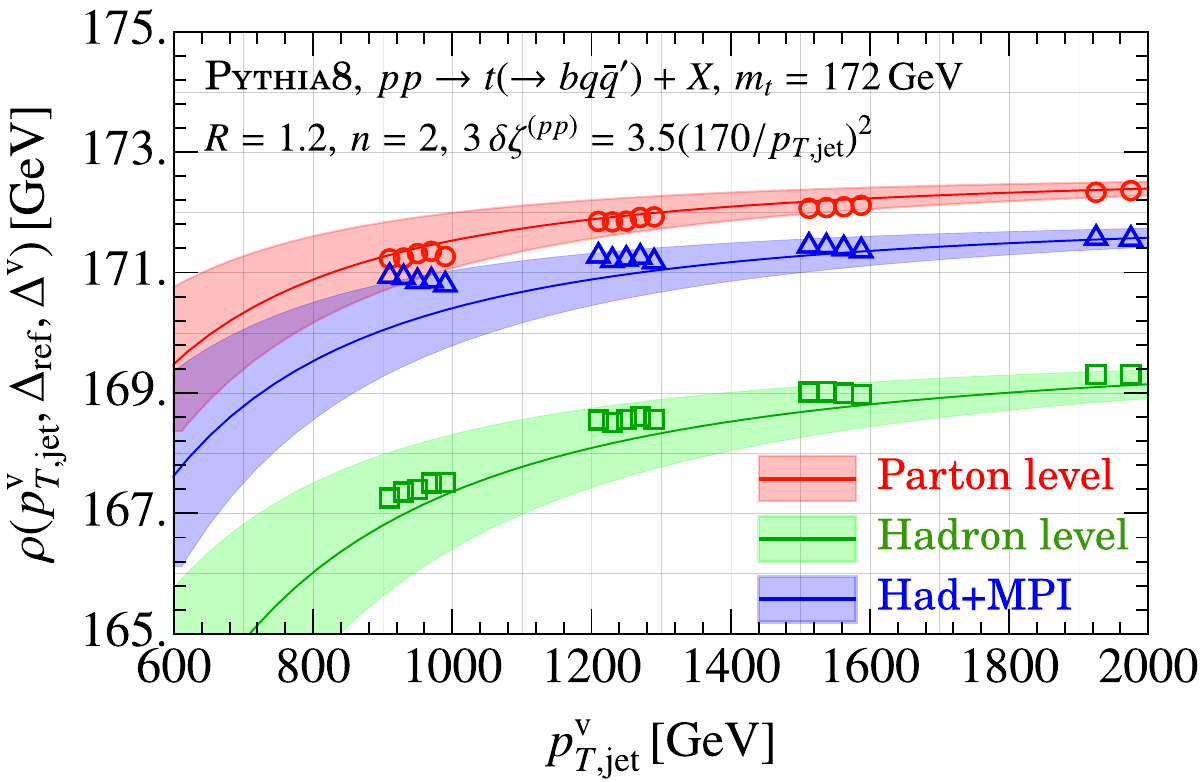}
	\caption{This figure shows parton shower data for $\rho$ generated in \Pythiaxx at parton level, hadron level, and with MPI (shown in open markers) overlaid with curves from \eq{rhoexpansion} demonstrating the self-consistency of our results. Five data sets for $\rho$ were averaged over with $p^{\text{ref}}_{T,\text{jet}} \in [550,650]$ GeV binned in $20$ GeV intervals.}
	\label{fig:yfit}
\end{figure}
Thus, we fit for $F_{\text{pert}}$ using the following procedure:
\begin{enumerate}
	\item Following \eq{rhoexpansion} we fit for the asymptote of $\rho$ (which we label $\rho_{\text{asy}}$) using a polynomial in $(1/p^{\text{v}}_{T,\text{jet}})^{n}$. In this paper we found that a third degree polynomial, 
	\begin{align}\label{eq:rhofit}
		\rho(p_{T,\rm jet}) = \rho_{\text{asy}} + c_{2}(p^{\text{v}}_{T,\text{jet}})^{-2} + c_{3}(p^{\text{v}}_{T,\text{jet}})^{-3}	\,	,	
	\end{align} 
	optimized the reduced $\chi^2$. The value of $\rho_{\text{asy}}$ was found to be stable, within our statistical accuracy, against the inclusion of further higher order terms, $c_{n}(p^{\text{v}}_{T,\text{jet}})^{-n}$. \Fig{asyFit} shows one such fit. No error bars are shown in this \Fig{asyFit} as it was produced from a single Monte Carlo sample. Fits of five samples are averaged over to produce the results and their errors in \Tab{resultsmt}.
	\item We extract $\Delta'^{\text{ref}}$ from the top-jet $p_{T}$ spectrum as shown in \Fig{pt_shifts}.
	\item Finally, we compute $F_{\text{pert}}$ using the asymptote of $\rho$, $\rho_{\rm asy}$, defined above in \eq{rhofit} as
	\begin{align}
		\sqrt{F_{\text{pert}}} = \rho_{\text{asy}} ~\frac{p^{\text{ref}}_{T, \mathrm{jet}}  + \Delta^{\text{ref}}}{p^{\text{ref}}_{T, \mathrm{jet}}}	\,	.
	\end{align} 
\end{enumerate}
The outcome of this procedure is given in \Tab{resultsmt} which shows the extracted $F_{\text{pert}}$ from \Pythiaxx with $m_{t} = 172$ GeV and $173$ GeV. The important outcome of this analysis is that the differences between the measured masses with parton, hadron and hadron+MPI data are $\lesssim 1$GeV and are smaller than the statistical errors. This analysis was not optimized to give a good statistical error and certainly can be improved. Thus we find promising evidence that complete theoretical control of the top mass, up to errors $<1$GeV, is possible with an EEEC based measurement.

To cross-check our result, purely to demonstrate self-consistency, in \Fig{yfit} we illustrate a theory fit of $\rho$ using parton shower data from \Pythiaxx with $m_{t} = 172$ GeV at parton level and hadron level.  The curves in \Fig{yfit} are \textit{not} the third degree polynomial used to extract $\rho_{\text{asy}}$ in \eq{rhofit}. Rather, the curves are, truncated at second order, using the values of $F_{\text{pert}}$ given in \Tab{resultsmt} and the values of $\Delta'^{\text{ref}}$ given in \Fig{pt_shifts}. Error bars correspond to the errors on $F_{\text{pert}}$ and $\Delta^{\text{ref}}$. 
To illustrate the partonic curve, a value of $\Delta^{\rm ref}_{\text{pert}}=(11\pm 3)$ GeV has been used which was extracted from the fit for $\rho^{(2)}$ (i.e. $c_{2}$ in \eq{rhofit}). This $\Delta^{\rm ref}_{\text{pert}}$ is not used in any of the preceding analysis (or anywhere else in this article) where all dependence on $\Delta^{\rm ref}_{\text{pert}}$ is absorbed in to the definition of $F_{\text{pert}}$. Each error band shows the combined statistical error from the determination of the asymptote and of $\Delta^{\rm ref}$ (including the dominant 3 GeV error on $\Delta^{\rm ref}_{\text{pert}}$). 

We find agreement between the MC data and our theory fit. \Fig{topmasshadron} along with \Fig{pp_MPI} further demonstrates the excellent agreement between theory and parton shower data wherein we fit $\zeta_{\text{peak}}(p_{t,\text{jet}})$ with the ansatz in \eq{pp_master}, also using the values for $F_{\text{pert}}$ in \Tab{resultsmt} and the values of $\Delta'^{\text{ref}}$ given in \Fig{pt_shifts}.

\begin{widetext}
	
	\begin{center}
		\begin{table}
			\begin{tabular}{||c|c|c|c||}
				\hline
				\Pythiaxx $m_{t}$ & EEEC Parton $\sqrt{F_{\text{pert}}}$ & EEEC Hadron $\sqrt{F_{\text{pert}}}$ & EEEC Hadron + MPI $\sqrt{F_{\text{pert}}}$ \\ \hline
				$172$ GeV & $172.6\pm 0.3$ GeV & $172.4 \pm 0.2 \pm 0.5$ GeV & $172.3 \pm 0.2 \pm 0.4$ GeV\\
				$173$ GeV & $173.5 \pm 0.3$ GeV & $173.9 \pm 0.3 \pm 0.5$ GeV & $173.6\pm 0.2 \pm 0.4$ GeV\\
				$175$ GeV & $175.5 \pm 0.4$ GeV & $175.2\pm 0.3 \pm 0.5$GeV & $175.1\pm 0.3 \pm 0.4$ GeV\\ \hline
				$173-172$ & $0.9 \pm 0.4$ GeV & $1.5 \pm 0.8 $ GeV & $1.3\pm 0.6 $ GeV\\ 
				$175-172$ & $2.9 \pm 0.5$ GeV & $2.8 \pm 0.8$ GeV & $2.8\pm 0.6 $ GeV\\
				$175-173$ & $2.0 \pm 0.5$ GeV & $1.3 \pm 0.8$ GeV & $1.5\pm 0.6 $ GeV\\\hline
			\end{tabular}
			\caption{A more detailed version of \Tab{pp_table}, showing separate results at parton, hadron and hadron+MPI level. Five data sets for $\rho$ were averaged over with $p^{\text{ref}}_{T,\text{jet}} \in [550,650]$GeV binned in $20$GeV intervals and $p^{\text{v}}_{T,\text{jet}} \in [900,2000]$GeV. One such data set and its fit is shown in \Fig{asyFit}. In each column the first error is from the fit of the $\rho$ asymptote and is statistical. The second error (when given) is also statistical and is the error from using the parton shower to determine $\Delta'^{\text{ref}}\approx \Delta^{\text{ref}}$ as extracted from the top jet $p_{t}$ distribution. Errors have been combined in quadrature in the final row. No theory errors are given.}
			\label{tab:resultsmt}
		\end{table}
	\end{center}

\begin{figure}[t!]
	\centering
	\includegraphics[width=0.45\textwidth]{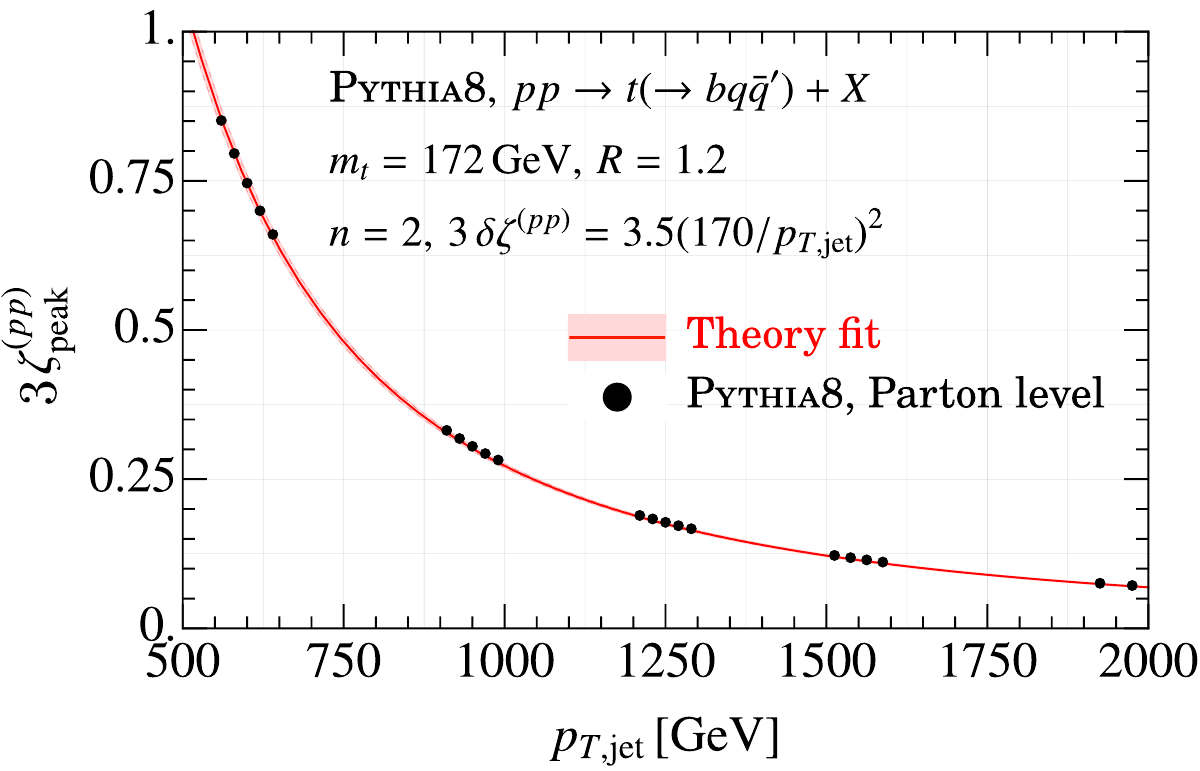}
	\includegraphics[width=0.45\textwidth]{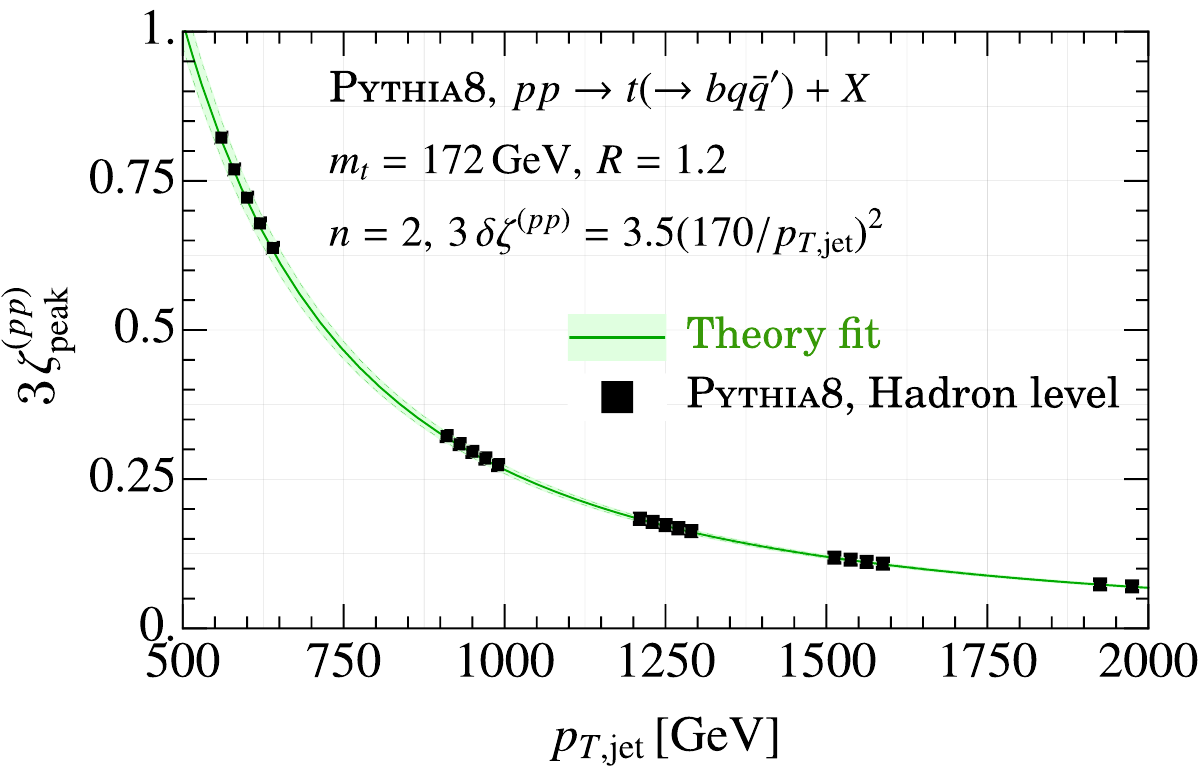}
	\caption{Energy correlator peak positions as a function of $p_{T,\text{jet}}$ at parton level and hadron level. (See \Fig{pp_MPI} for an analogous figure including MPI). The theory fit uses the ansatz in \eq{pp_master} with the values of $F_{\text{pert}}$ given in \Tab{resultsmt} and $\Delta'^{\text{ref}}$ extracted in \Fig{pt_shifts}. Excellent agreement between the theoretical fit and \Pythiaxx is observed in all cases.}
	\label{fig:topmasshadron}
\end{figure}

\end{widetext}

\bibliographystyle{apsrev4-1}

\bibliography{spinning_gluon}

\newpage
\clearpage

\end{document}

%% file: commands.tex
\usepackage{bm}
\usepackage{graphicx}
\usepackage{amsmath,amssymb,mathtools} 
\usepackage{xspace}
\usepackage{wrapfig}
\usepackage{slashed}
\usepackage[normalem]{ulem}
\usepackage[dvipsnames]{xcolor}
\usepackage[utf8]{inputenc}
\usepackage[T1]{fontenc}
\usepackage{mathtools}
\usepackage{stmaryrd}
\usepackage{appendix}
\usepackage{xspace}
\usepackage{upgreek}
\usepackage{multirow}

\usepackage{scrextend}
\usepackage[colorlinks=true
,urlcolor=OliveGreen
,anchorcolor=OliveGreen
,citecolor=OliveGreen
,filecolor=OliveGreen
,linkcolor=OliveGreen
,menucolor=OliveGreen
,linktocpage=true
,pdfa=true]{hyperref} 

\usepackage{tabularx}
\usepackage{tensor}

\usepackage{marginnote}

\definecolor{darkgreen}{rgb}{0.0, 0.4, 0.0}

\newcommand{\df}{\mathrm{d}}

\newcommand{\ra}{\rightarrow}

\newcommand{\as}{\alpha_s}

\newcommand{\cN}{{\mathcal N}}

\newcommand{\nn}{\nonumber}

\newcommand{\Pythiaxx}{\texttt{Pythia\xspace8.3}\xspace}

\newcommand{\Vinciaxx}{\texttt{Vincia\xspace2.3}\xspace}

\newcommand{\Fastjet}{\texttt{FastJet}\xspace}

\def\ln{\textrm{ln}}
\def\df{\textrm{d}}
\def\nn{\nonumber}

\def\LQCD{\Lambda_{\rm QCD}}

\newcommand{\ee}{{e^+e^-}}

\allowdisplaybreaks[2]

\newcommand{\cO}{\mathcal{O}}

\newcommand{\cM}{\mathcal{M}}
\newcommand{\cE}{\mathcal{E}}

\DeclareRobustCommand{\Refcite}[1]{Ref.~\cite{#1}}

\DeclareRobustCommand{\eq}[1]{eq.~\eqref{eq:#1}}

\DeclareRobustCommand{\secn}[1]{\hyperref[sec:#1]{section~\ref*{sec:#1}}}
\DeclareRobustCommand{\Sec}[1]{\hyperref[sec:#1]{Section~\ref*{sec:#1}}}
\DeclareRobustCommand{\secs}[2]{sections~\ref{sec:#1} and \ref{sec:#2}}

\DeclareRobustCommand{\subsec}[1]{\hyperref[subsec:#1]{subsection~\ref*{subsec:#1}}}
\DeclareRobustCommand{\Subsec}[1]{\hyperref[subsec:#1]{Subsection~\ref*{subsec:#1}}}

\DeclareRobustCommand{\app}[1]{\hyperref[app:#1]{appendix~\ref*{app:#1}}}
\DeclareRobustCommand{\App}[1]{\hyperref[app:#1]{Appendix~\ref*{app:#1}}}

\DeclareRobustCommand{\fig}[1]{\hyperref[fig:#1]{figure~\ref*{fig:#1}}}
\DeclareRobustCommand{\Fig}[1]{\hyperref[fig:#1]{Figure~\ref*{fig:#1}}}

\DeclareRobustCommand{\tab}[1]{\hyperref[tab:#1]{table~\ref*{tab:#1}}}
\DeclareRobustCommand{\Tab}[1]{\hyperref[tab:#1]{Table~\ref*{tab:#1}}}